\documentclass[twocolumn,notitlepage,nofootinbib,floatfix]{revtex4-1}

\pdfoutput=1

\usepackage{amsmath}
\usepackage{amsfonts}
\usepackage{amssymb}
\usepackage{mathrsfs}
\usepackage{amsthm}
\usepackage{wasysym}
\usepackage[usenames]{color}
\usepackage[usenames]{xcolor}
\usepackage[unicode,colorlinks=true,linkcolor=blue,citecolor=gray,pdfusetitle]{hyperref}
\usepackage[all]{hypcap}
\usepackage{graphicx}
\usepackage{xspace}
\usepackage{verbatim}
\usepackage{enumerate}

\usepackage[T1]{fontenc}
\usepackage[utf8]{inputenc}

\usepackage{microtype}

\usepackage{multirow}

\newcommand{\nn}{\nonumber}

\newcommand{\cd}{\nabla}

\newcommand{\lie}{\mathcal{L}}

\newcommand{\dvol}{\mathrm{dVol}\,}

\newcommand{\SLTR}{\ensuremath{SL(2,\mathbb{R})}}
\newcommand{\SLTRU}{\ensuremath{SL(2,\mathbb{R})\times U(1)}}
\newcommand{\sltr}{\ensuremath{\mathfrak{sl}(2,\mathbb{R})}}
\newcommand{\sltru}{\ensuremath{\mathfrak{sl}(2,\mathbb{R})\times \mathfrak{u}(1)}}

\begin{document}

\title{Separating metric perturbations in near-horizon extremal Kerr}

  \author{Baoyi Chen}
  \email{baoyi@tapir.caltech.edu}
  \author{Leo C.~Stein}
  \email{leostein@tapir.caltech.edu}
  \affiliation{%
TAPIR,
Walter Burke Institute for Theoretical Physics,
California Institute of Technology, Pasadena, CA
91125, USA}

\hypersetup{pdfauthor={Chen and Stein}}

  \date{\today}

  \begin{abstract}
    Linear perturbation theory is a powerful toolkit for studying
    black hole spacetimes.
    However, the perturbation equations are hard to solve unless we
    can use separation of variables.
    In the Kerr spacetime, metric perturbations do not separate, but
    curvature perturbations do.  The cost of curvature perturbations
    is a very complicated metric-reconstruction procedure.
    This procedure can be avoided using a symmetry-adapted choice of
    basis functions in highly symmetric spacetimes, such as
    near-horizon extremal Kerr.
    In this paper, we focus on this spacetime, and
    (i) construct the symmetry-adapted basis functions;
    (ii) show their orthogonality; and
    (iii) show that they lead to separation of variables of the
    scalar, Maxwell, and metric perturbation equations.
    This separation turns the system of partial differential equations
    into one of ordinary differential equations over a compact domain,
    the polar angle.
  \end{abstract}

 \maketitle

\section{Introduction}
\label{sec:introduction}

Linear metric perturbation theory is widely used in studying weakly-coupled
gravity~\cite{Wald}.  For example, it can be applied to investigating
the stability of black holes, gravitational radiation produced by
material sources moving in a curved background, and so on.  In the
context of linearized gravity, the equations that describe
gravitational perturbations are the linearized Einstein equations
(LEE).  Although they are linear, the LEE are still difficult to solve
unless we can separate variables.  In the Kerr spacetime, while in
Boyer-Lindquist (BL) coordinates $t$ and $\phi$ can be separated, $r$
and $\theta$ remain coupled due to lack of
symmetry~\cite{Teukolsky:2014vca}.

A successful approach towards separating wave equations for
perturbations of the Kerr black hole was first developed by
Teukolsky~\cite{Teukolsky:1972my,Teukolsky:1973ha}.  Instead of
looking at metric perturbations, Teukolsky adopted the Newman-Penrose
(NP) formalism~\cite{Newman:1961qr} and obtained a separable wave
equation for Weyl curvature tensor components $\Psi_0$ and $\Psi_4$.
The spin-weighted version of this equation, known as the Teukolsky
equation, not only works for gravitational perturbations, i.e.~tensor
fields, but can also be applied to scalar, vector, and spinor fields.
To obtain the other Weyl scalars and recover the perturbed metric, one
has to go through a complicated metric reconstruction procedure.  The
methods were independently developed by
Chrzanowski~\cite{Chrzanowski:1975wv} and by Cohen and
Kegeles~\cite{Kegeles:1979an}, in which they obtain the perturbed
metric via an analogue of Hertz potentials.
However, these methods only apply to certain gauge choices and
vacuum or highly-restricted source terms~\cite{Whiting:2005hr}.

The desire for separable equations, the complication of metric
reconstruction along with gauge- and source-restrictions, motivate us
to try to develop a new formalism for studying metric perturbations in
the Kerr spacetime, in a covariant, gauge-invariant way.

The metric perturbation equation may not be separable in Kerr, but
Schwarzschild perturbations have long been known as separable due to
the time translation invariance and spherical
symmetry~\cite{Regge:1957td, Vishveshwara:1970cc, Zerilli:1970se,
  Zerilli:1971wd}.  The gauge-independent language of Schwarzschild
perturbations was started by Sarbach and Tiglio~\cite{Sarbach:2001qq},
and brought to fruition by Martel and Poisson~\cite{Martel:2005ir}.
In the Schwarzschild background, metric perturbations are expanded in
scalar, vector, and symmetric tensor spherical harmonics.  These basis
functions naturally lead to separation of variables in the LEE.

Schematically, the separation of variables in some differential
equations of motion, such as the scalar wave equation, Maxwell's
equations, and the linearized Einstein equations, can all be
understood via
\begin{align}
  \mathcal{D}_{x}\!
  \Big[
  \Big(
  \parbox{1.1cm}{\centering\tiny symmetry \\ adapted \\ basis}
  \Big)
  \!\times\!
  \Big(
  \parbox{1.2cm}{\centering\tiny dependence \\ on rest of \\ coordinates}
  \Big)
  \Big]
  \!=\!
  \Big(
  \parbox{1.1cm}{\centering\tiny symmetry \\ adapted \\ basis}
  \Big)
  \!\times\!
  \mathcal{D}_{x^\prime}\!
  \Big[
  \parbox{1.2cm}{\centering\tiny dependence \\ on rest of \\ coordinates}
  \Big]
  \nn
  .
\end{align}
Here $\mathcal{D}_x[\cdot]$ is some isometry-equivariant differential
operator.  If the argument is decomposed in a natural isometry-adapted
basis, then these basis functions pull straight through the
differential operator, leaving new operators $\mathcal{D}_{x'}[\cdot]$
which only act on the remaining non-symmetry coordinates.

We show that this type of reduction is true for a special limit of
Kerr spacetime: the near-horizon extremal Kerr (NHEK).  This
spacetime was introduced in~\cite{Bardeen:1999px} as an analogue of
$AdS_2\times S^2$.  The NHEK limit exhibits a symmetry group that is
``enhanced'' relative to Kerr: the spacetime has four Killing
vector fields that
generate the isometry group \SLTRU.  The three-dimensional orbit space
of the isometry reduces the system of partial differential equations (PDEs)
to one of ordinary differential equations (ODEs), leading to separable
equations of motion.  This is achieved by expanding unknown tensors
into some basis functions adapted to the isometry.  In this paper, we
(i)~construct these basis functions, (ii)~prove orthogonality in
geodesically-complete coordinates, and (iii)~show separation of
variables in the differential equations for some physical systems.
With these accomplishments, we arrive at a new formalism to deal with
(extremal) Kerr perturbation that differs from using metric
reconstruction on solutions to the Teukolsky equation.  In this
formalism there will be no gauge preference, no complications of
solving PDEs, but rather only ODEs.
This greatly reduces the amount of work while studying perturbations
of extremal Kerr black holes, whether in GR or beyond-GR theories.

We organize the paper as follows.
In Sec.~\ref{sec:kerr-nhek-limit} we review the NHEK limit of the Kerr
black hole, and elaborate on the structure of NHEK's isometry Lie
group $\SLTRU$.
In Sec.~\ref{sec:high-lowest-weight}, we construct the highest-weight
module for NHEK's isometry group, and obtain the
scalar/vector/symmetric tensor basis functions.
In Sec.~\ref{sec:orthogonality-basis} we present a proof of
orthogonality for the basis functions in global coordinates.
In Sec.~\ref{sec:separation-variables} we show that with these bases,
we can separate variables in the scalar Laplacian, Maxwell system, and
linearized Einstein equation.
Finally we conclude and discuss future work in
Sec.~\ref{sec:concl-future-work}.

\section{Kerr and the NHEK limit}
\label{sec:kerr-nhek-limit}
In this paper we choose geometric units $(G = c = 1)$ and signature
$(-{}+{}+{}+)$ for our metric $g$ on the spacetime manifold
$\mathcal{M}$.
A rotating, asymptotically-flat black hole in vacuum general
relativity is described by the Kerr metric~\cite{Kerr:1963ud}.  For
simplicity we will set the mass to $M=1$.  In BL
coordinates $(t, r, \theta, \phi)$ the line element of the Kerr black
hole is given by~\cite{Boyer:1966qh}
\begin{align}
  \text{d}s^2 = 
  &- \frac{\Delta}{\Sigma} (\text{d}t - a\sin^2\theta\,\text{d}\phi)^2 
  + \frac{\Sigma}{\Delta}\,\text{d}r^2 
  + \Sigma\,\text{d}\theta^2 \\ \nonumber
  &+ \frac{\sin^2\theta}{\Sigma} 
  \left[ (r^2 + a^2)\,\text{d}\phi - a\,\text{d}t\right]^2,
\end{align}
where $\Delta = r^2 - 2r + a^2$ and $\Sigma = r^2 + a^2 \cos^2\theta$.
The ranges of the BL coordinates are given by
$t\in(-\infty, +\infty)$, $r\in(0,+\infty)$, $\theta\in[0,\pi]$,
$\phi\in[0,2\pi)$.  In this paper we focus on a particular scaling limit
of Kerr.  This limit is usually described by
the scaling coordinates $(T,\Phi,R)$ introduced
in~\cite{Bardeen:1999px}, which are related to the BL coordinates via
\begin{align}
  t &= \frac{2 T}{\lambda}\,, &
  \phi &= \Phi + \frac{T}{\lambda}\,, &
  r &= 1 + \lambda R\,.
\end{align}
We also introduce a new coordinate $u$ for the polar angle via
$u=\cos\theta$.  The NHEK limit is then obtained by taking the
$(a\to M, \lambda \to 0)$ limit of the Kerr metric in these
coordinates, which yields the line element
\begin{align}
  \text{d}s^2 = 
  2 \Gamma(u)\bigg[ &-R^2\,\text{d}T^2 + \frac{\text{d}R^2}{R^2}
  + \frac{\text{d}u^2}{1 - u^2} \\ \nonumber
  &+ \Lambda(u)^2(\text{d}\Phi + R\,\text{d}T)^2 \bigg], 
\end{align}
where $\Gamma(u) = (1 + u^2)/2$ and
$\Lambda(u) = 2 \sqrt{1-u^2}/ (1 + u^2)$.  This metric is interpreted
on the region $T\in(-\infty,+\infty)$, $\Phi\in[0,2\pi)$,
$R\in(0,+\infty)$, $u\in[-1,1]$.

From now on we will refer to $(T,\Phi,R,u)$ as \textit{Poincar\'{e}
  coordinates}. The $T,R$-coordinates of NHEK are similar to the
Poincar\'{e} coordinates on the two-dimensional anti-de Sitter space
$AdS_2$, which only cover a subspace of the global spacetime called
the \textit{Poincar\'{e} patch}. In particular, the $u=\pm1$
submanifolds are both precisely $AdS_2$.  We can make this metric
geodesically complete by defining the \textit{global coordinates}
$(\tau,\varphi,\psi,u)$ according to~\cite{Bardeen:1999px}
\begin{align}
T &=\frac{\sin\tau}{\cos\tau - \cos \psi}, \quad
R = \frac{\cos\tau - \cos \psi}{\sin \psi},\\ \nonumber
\Phi &= \varphi + \ln\left|\frac{\cos\tau - \sin\tau \cot\psi}{1+\sin\tau \csc\psi}\right|,
\end{align}
where $\tau \in (-\infty,+\infty)$, $\psi \in [0, \pi]$, 
$\varphi \sim \varphi + 2\pi$.
The NHEK metric in global coordinates is
\begin{align}
  \text{d}s^2 = 2 \Gamma(u)\bigg[& (-\text{d}\tau^2 + \text{d}\psi^2)\csc^2\psi 
               + \frac{\text{d}u^2}{1 - u^2} + \\ \nonumber
               &+\Lambda(u)^2(\text{d}\varphi - \cot\psi\,\text{d}\tau)^2 \bigg].
\end{align}

The NHEK spacetime has four Killing vector fields (KVFs), which
generate the isometry group $G\equiv \SLTRU$.
The four generators in Poincar\'{e} coordinates are given by
\begin{align}
  &H_0\, = T\,\partial_T - R\,\partial_R, \\ \nonumber
  &H_+   = \partial_T,   \\ \nonumber
  &H_-   = (T^2 + \frac{1}{R^2})\,\partial_T - 2\,TR\,\partial_R 
           - \frac{2}{R}\,\partial_\Phi, \\ \nonumber
  &Q_0\,\, = \partial_\Phi.
\end{align}
$H_0$ is the infinitesimal generator of \textit{dilation}, which
leaves the metric invariant under $R \rightarrow cR$ and
$T \rightarrow T/c$ for some constant $c\in(0,+\infty)$. $Q_0$ is the
generator of the rotation along $\Phi$ which generates the $U(1)$
group. $H_+$ is the time translation generator inherited from Kerr.
The four generators form a \emph{representation} $\rho_{P}$ of the Lie
algebra
$\mathfrak{g} \equiv \sltru $,
\begin{align}
  \label{eq:Lie-algebra-poincare}
  [H_0 \,, H_\pm] &= \mp H_\pm \,, \\ \nonumber
  [H_+ \,, H_-] &= 2\,H_0 \,,      \\ \nonumber
  [H_s \,, Q_0] &= 0 \,. \qquad (s=0,\pm)
\end{align}

In global coordinates, we can similarly obtain four (different) generators
that are KVFs of the NHEK spacetime,
\begin{align}
  L_\pm &= i e^{\pm i \tau} \sin\psi (-\cot\psi\partial_\tau \mp 
           i\partial_\psi + \partial_\varphi), \\ \nonumber
  L_0 &= i \partial_\tau,   \\ \nonumber
  W_0 &= -i \partial_{\varphi}.
\end{align}
This is a different representation, $\rho_{g}$.  But since it is still
a Lie algebra representation, they satisfy the same commutation
relations as in Eq.~\eqref{eq:Lie-algebra-poincare} with all $H$'s
replaced by $L$'s, and $Q_0$ replaced $W_0$.

We say that the group $G$ acts on the manifold $\mathcal{M}$ by
translation, $G \circlearrowleft \mathcal{M}$.  That is, every element
$g\in G$ determines an isomorphism $\phi_{g}: \mathcal{M} \to
\mathcal{M}$, and these isomorphisms, under composition, form a
representation of the group $G$.  There is an induced action on the
space of functions/vector fields/forms/tensors/etc.~living on
$\mathcal{M}$ by pullback under the map $\phi_{g}$~\cite{MR2954043}.  We call the
pullback $\phi^{*}_{g}$, overloading this symbol to mean the pullback
from sections of any tensor bundle to itself.  In this way, the group
also acts on all spaces of $(p,q)$-tensors.

Studying the neighborhood of the identity $e\in G$, we get the induced
action of the Lie algebra $\mathfrak{g}$ on these same tensor bundles.
The infinitesimal version of a pullback of a tensor field is the Lie
derivative of that field~\cite{MR2954043}.  Thus the induced
action of $\mathfrak{g}$
on tensors is Lie derivation along the representation of the Lie
algebra element.  That is, given a representation as tangent vector
fields $\rho: \mathfrak{g} \to \mathfrak{X}(\mathcal{M})$, for some
algebra element $\alpha\in \mathfrak{g}$, the induced action of $\alpha$ on a
tensor $\mathbf{t}$ is via the Lie derivative,
\begin{align}
  \alpha \cdot \mathbf{t} = \lie_{\rho(\alpha)} \mathbf{t} \,.
\end{align}

One of the crucial algebra elements we will need is the Casimir
element of the $\sltr$ factor.  Let $h_{0}, h_{\pm} \in
\mathfrak{g}$ be the algebra elements whose representations are
$\rho_{P}(h_{s})=H_{s}$ for $s = 0, \pm$.  Then the Casimir element
of the $\sltr$ factor, in this basis, is proportional to
\begin{align}
  \label{eq:Casimir-def}
  \Omega \equiv h_{0} (h_{0} - 1) - h_{-} h_{+} \,,
\end{align}
which commutes with every element of $\mathfrak{g}$.  Under the
Poincar\'e-coordinates representation $\rho_{P}$, the Casimir acts on
tensors via
\begin{align}
  \label{eq:Casimir-lies}
  \Omega \cdot \mathbf{t} =
  \left(
  \lie_{H_{0}} ( \lie_{H_{0}} - \text{id} )
  - \lie_{H_{-}} \lie_{H_{+}}
  \right) \mathbf{t} \,.
\end{align}
By construction, the differential operator on
the right-hand side of Eq.~\eqref{eq:Casimir-lies} commutes with
$\lie_{X}$, where $X$ is one of $\{ H_{0}, H_{\pm}, Q_{0} \}$.
Similarly, under the global-coordinates representation $\rho_{g}$, the
Casimir acts as in Eq.~\eqref{eq:Casimir-lies}, but with $H$'s
replaced with $L$'s; and this operator will similarly commute with
$\lie_{X}$ where $X$ is one of $\{ L_{0}, L_{\pm}, W_{0} \}$.

\section{The highest- (lowest-)\break{} weight method}
\label{sec:high-lowest-weight}

In this section we construct the scalar, vector, and symmetric tensor
bases for NHEK's isometry group $\SLTRU$. First we briefly review the
formalism of finding basis functions adapted to the isometry group in
Schwarzschild spacetime.  By drawing analogy to the Schwarzschild case
and further utilizing the \textit{highest- (lowest-)weight method} for
non-compact groups, we will be able to construct unitary
representations of NHEK's isometry group.

\subsection{Review: Unitary representations\break{} of $SO(3)$ in Schwarzschild}
\label{sec:uni-reps-iso-Sch}

The full spacetime manifold of Schwarzschild spacetime is
$\mathcal{M}_{\mathrm{Sch}}=M^2\times S^2$.  The two-dimensional
submanifold $M^2$ is the $(\bar{t},\bar{r})$-plane, and $S^2$ is the
unit two-sphere coordinated by $(\bar{\theta},\bar{\phi})$.  Here
$(\bar{t},\bar{r},\bar{\theta},\bar{\phi})$ are the usual
Schwarzschild coordinates.  Part of the isometry group of
Schwarzschild is $SO(3)$, which acts on the $S^{2}$ factors.  The
three generators of the group are simply the rotations along each
Cartesian axis, i.e.~$J_x, J_y, J_z \in \mathfrak{so}(3)$.  The
Casimir operator of $\mathfrak{so}(3)$ is given by
$J^2 = J^2_x+J^2_y+J^2_z$.

In any space that $SO(3)$ acts upon, we can look for bases of
functions which simultaneously diagonalize $J^{2}$ and $J_{z}$---that
is, they are eigenfunctions of both operators.  In the space of
complex functions on the unit sphere, these eigenfunctions turn out to
be the spherical harmonic functions $Y^{\mu,\nu}$, where $\mu,\nu$
label the functions (they are not tensor indices).  The even/odd
parity vector harmonics, $Y_{A}^{\mu,\nu}, X_{A}^{\mu,\nu}$, and
tensor harmonics, $Y_{AB}^{\mu,\nu}, X_{AB}^{\mu,\nu}$, are also
simultaneous eigenfunctions of $J^{2}$ and $J_{z}$ (where now $A,B$
are (co-)tangent indices on $S^{2}$).  All of the
scalars, vectors, and tensors here have eigenvalue $-\mu(\mu+1)$ for
the operator $J^{2}$, and eigenvalue $i\nu$ for $J_{z}$.

Under any rotation, scalar spherical harmonics with different values
of $\mu$ may not rotate into each other.  In this sense, the function
space has been split up into diagonal blocks labeled by $\mu$.  We say
that each $\mu$ block ``lives in'' or ``transforms under'' a
representation of $SO(3)$.

We have not yet imposed regularity or tried to make these
representations unitary.  Let us define the raising and lowering
operators $J_\pm=J_x \pm i J_y$, which increase/decrease the $\nu$
index (eigenvalue of $-i J_{z}$) by one.  A \emph{highest-weight}
state is one which is annihilated by the raising operator,
$J_{+} f = 0$, and similarly a lowest-weight state is annihilated by
the lowering operator.  For spherical harmonics, we find that the
highest-weight condition imposes that $\nu=\mu$, and $Y^{\mu,\mu}$ is
annihilated by $J_{+}$.  Similarly, the lowest-weight condition
imposes that $\nu = -\mu$.

From the representation theory of compact simple Lie groups, irreducible
unitary representations must be finite-dimensional~\cite{Barut:1986dd}.
Therefore, if we start with a highest-weight state $Y^{\mu,\mu}$,
after a finite number of applications of the lowering operator, we
must end on a lowest-weight state $Y^{\mu,-\mu}$.  This gives us the
condition that $2\mu+1$ is a positive integer, or $\mu=0, \frac{1}{2},
1, \ldots$.  Periodicity in the azimuthal angle $\bar{\phi}$ gives the
condition that $\nu$ must be an integer $m$.  This gives the ordinary
spherical harmonics $Y^{l,m}$.  The same arguments apply to the
vector and tensor representations.

Since these bases are adapted to the isometry group of Schwarzschild,
they readily lead to a separation of variables in the linearized
Einstein equations~\cite{Martel:2005ir}.

\subsection{Unitary representations\break{} of $\SLTRU$ in NHEK}
\label{sec:uni-reps-iso-nhek}

We now apply the highest-/lowest-weight formalism to NHEK.  In the
Schwarzschild spacetime, the orbit space of the isometry $SO(3)$ is
$S^2$, therefore we expect a $2+2$ decomposition of the whole
manifold.  Similarly, in the NHEK spacetime, the isometry group
$\SLTR\times\nobreak{}U(1)$ acts on the three-dimensional hypersurfaces $\Sigma_u$ of
constant polar angle $\theta$ (or $u$).  This enables us to perform a
$3+\nobreak 1$ decomposition of the spacetime.  In both cases, we can
simultaneously diagonalize some algebra elements, including the
Casimir, in various tensor spaces.

However there is an important difference between the two spacetimes.
In the NHEK case, we encounter the non-compact group $\SLTR$.  It is
known that for non-compact simple Lie groups like $\SLTR$, the only
irreducible unitary finite-dimensional representation is the trivial
representation~\cite{Barut:1986dd}.  As a result, one can find two
distinct unitary representations of $\SLTRU$: the
\textit{highest-weight module} or the \textit{lowest-weight
  module}. Both of them
are infinite-dimensional representations in the NHEK case. For compact
groups like $SO(3)$, these two modules coincide.

Our method to find the general (scalar, vector, and symmetric tensor)
basis functions $\xi$ associated with the highest-weight module of
NHEK's isometry can be summarized into four steps. Notice that the
method presented here is not restricted to NHEK spacetime.  For
instance it can also be applied to finding the basis functions in
near-horizon near-extremal Kerr (near-NHEK) which has the same
isometry group as NHEK's~\cite{Bredberg:2009pv}.
This will be left for future work.  For
readers who are more interested in what the bases of NHEK's isometry
look like either in Poincar\'{e} or global coordinates, the explicit
expressions are given in App.~\ref{app:S-V-T-Basis}.

\paragraph{Orbit space.}
For each point $p\in\mathcal{M}$, there is the orbit
$Gp = \{ \phi_{g}(p) | g\in G\}$, all points which are related to $p$
by an $\SLTRU$ transformation.  $Gp$ is a 3-dimensional submanifold of
$\mathcal{M}$, and the collection of all the orbit spaces forms a
foliation.  In this case, each leaf $\Sigma_{u}$ is a surface of
constant $\theta$ (or $u$).  Thus we can perform a $3+1$ decomposition
of the spacetime, and look for basis functions of $\SLTRU$ acting on a
hypersurface $\Sigma_u$.

\paragraph{Highest weight states.}
\label{sec:step-highest-weight}
Second, we simultaneously diagonalize
$\{\lie_{Q_{0}}, \lie_{H_0}, \Omega\}$ in the space of scalar, vector,
and symmetric tensor functions. We label the eigenstates by $m,h,k$
respectively,
\begin{align}
  \label{eq:codiagonalize-three-operator}
  \lie_{Q_0}\,{\xi}^{(m\,h\,k)} &= im\,{\xi}^{(m\,h\,k)} \,, \\
  \Omega\,{\xi}^{(m\,h\,k)}    &= h(h+1)\,{\xi}^{(m\,h\,k)} \,, \nn \\
  \lie_{H_0}\,{\xi}^{(m\,h\,k)} &= (-h+k)\,{\xi}^{(m\,h\,k)} \,. \nn
\end{align}
Then using the raising operator $\lie_{H_{+}}$,
we also impose the highest-weight condition, $k=0$,
\begin{equation}
  \label{eq:general-highest-weight-solution}
  \mathcal{L}_{H_+}\,{\xi}^{(m\,h\,0)} = 0 \,.
\end{equation}
The solutions ${\xi}^{(m\,h\,0)}$ that satisfy both
Eq.~\eqref{eq:codiagonalize-three-operator} and
\eqref{eq:general-highest-weight-solution} are the highest-weight
basis functions.  At each point on $\Sigma_{u}$, the spaces of scalars,
vectors, and symmetric tensors have dimensions 1, 3, and 6.  Thus the
space of solutions of this system of equations is a linear vector
space of dimension 1, 3, and 6 for scalars, vectors, and symmetric
tensors, for each choice of $(m, h)$.  Correspondingly, for each $(m,
h)$, there will be 1, 3, and 6 free coefficients $c_{\beta}$ for the
solution, with $\beta$ ranging over the appropriate dimensionality.

\paragraph{Descendants.}
\label{sec:step-descendants}
Next, we obtain basis functions with arbitrary weight by applying the
lowering operator $\mathcal{L}_{H_-}$ to the highest-weight states $k$
times, i.e.
\begin{equation}
  {\xi}^{(m\,h\,k)} = (\mathcal{L}_{H_-})^k\,{\xi}^{(m\,h\,0)}.
\end{equation}

\paragraph{Lifting to the whole manifold.}
Finally, we promote the basis functions living on $\Sigma_u$ to
functions living on the whole manifold $\mathcal{M}$ by sending all
unknown constant coefficients $c_{\beta}$ (from the end of step b) to
be unknown smooth functions $c_{\beta}(u)$.
While lifting the vector and tensor bases from $\Sigma_u$ to $\mathcal{M}$,
i.e.~$ V_i\rightarrow V_a $ and $W_{ij}\rightarrow W_{ab}$, we also 
set all their projections on the $u$ direction to be zero,
i.e.~$V_u=0$, $W_{iu}=W_{ui}=W_{uu}=0$.

To obtain the basis functions in global coordinates, one just replaces
$H_s$ by $L_s$, where $s=0,\pm$, and $Q_0$ by $iW_0$ in steps b and c.
To construct the lowest-weight modules of NHEK's isometry group, one
should instead impose the lowest-weight condition
$\lie_{H_-}\,{\xi}^{(m\,h\,0)} = 0$, and the condition
$\Omega\,{\xi}^{(m\,h\,k)} = h(h-1)\,{\xi}^{(m\,h\,k)}$,
in step b.  All descendant states
will then be obtained by applying the raising operator $\lie_{H_+}$ on
the lowest-weight states.  In Poincar\'{e} coordinates, we focus on
the basis functions that form the highest-weight module because their
expressions are simpler.  In global coordinates, we show both
representations explicitly in App.~\ref{app:scalar-basis-highest-reps}
and \ref{app:scalar-basis-lowest-reps}.  Unless otherwise specified,
our basis functions will refer to those obtained using the
highest-weight method.

Let us remark on the allowed values of $m,\,h,\,k$.  It is
straightforward to see $k\in \mathbb{Z}^{+}$ by construction,
and $m\in \mathbb{Z}$ due to the periodic boundary conditions for the
azimuthal angle.  In order to have a unitary representation of the
isometry group, there are conditions on $h$ as well.  For the scalar
case, for instance, if we apply the raising operator on a scalar in
the highest-weight module, we get
\begin{equation}
  \lie_{H_+}\,F^{(m\,h\,k)} = k(k-1-2h)\,F^{(m\,h\,k-1)}.
\end{equation}
A nontrivial unitary representation of NHEK's isometry group then
requires $k-1-2h \neq 0$, otherwise there would be a lowest-weight
state that would lead to a finite-dimensional (and hence non-unitary)
representation.  The same conclusion holds for either the vector or
the tensor bases.  The values of $h$ also depend on the regularity
conditions we impose.  For instance, in global coordinates, the
highest-weight scalar basis is proportional to
\begin{equation}
  F^{(m\,h\,0)} \propto  (\sin\psi)^{-h}  \exp[i (h \tau + m \varphi) +m \psi].
\end{equation}
Regularity at the boundaries $\psi=0$ and $\psi=\pi$ requires
$h \le 0$.  Another example is given in Sec.~\ref{sec:sep-scalar} when
we solve for the free massless scalar wave equation in the NHEK
spacetime, where $h$ must take on some fixed values due to the
regularity conditions for spheroidal harmonics.

\section{Orthogonality in\break{} global coordinates}
\label{sec:orthogonality-basis}

In this section we present a proof that all the scalar, vector, and
symmetric tensor basis functions of NHEK's isometry group, when given
in global coordinates, form orthogonal basis sets.  In this proof we
will use the vector basis functions defined on $\Sigma_u$ as an
example.  That is, they are functions of $\tau,\varphi,\psi$.  As we
shall see, lifting to the whole manifold $\mathcal{M}$ and extending
the proof to the scalar and symmetric tensor cases will be
straightforward.

Let us introduce the metric induced on the hypersurface $\Sigma_u$ as
$\gamma_{ij}$, and $D$ is the unique torsion-free Levi-Civita
connection that is compatible with $\gamma$.  Here Latin letters in
the middle of the alphabet ($i,j,k$) denote 3-dimensional tangent
indices on $\Sigma_{u}$.  Consider the vector
basis function ${\bf u}^{(m\,h\,k)}(\tau,\varphi,\psi)$ and
${\bf v}^{(m^\prime\,h^\prime\,k^\prime)}(\tau,\varphi,\psi)$. We
would like to show bases with different $m,h,k$ are orthogonal,
\begin{equation}
\langle {\bf u}, {\bf v} \rangle \equiv  \int_{\Sigma_u} \dvol
\overline{u_i^{(m\,h\,k)}} v^i_{(m^\prime\,h^\prime\,k^\prime)}
\propto \delta_{m,m^\prime} \delta_{h,h^\prime} \delta_{k,k^\prime}
\,.
\label{eq:orthogonal-relation}
\end{equation}
Here the overbar denotes complex conjugation, and the volume element
is given by
\begin{equation}
\int_{\Sigma_u} \dvol = \lim_{T\rightarrow \infty}
\int_{-T}^T\text{d}\tau \int_0^{2\pi}\text{d}\varphi
\int_0^{\pi}\text{d}\psi \sqrt{-\gamma}
\,,
\end{equation}
where $\gamma$ is the determinant of the three-dimensional metric,
and in these coordinates $\sqrt{-\gamma} =
2\csc^{2}\psi\sqrt{1-u^{4}}$.
To prove Eq.~\eqref{eq:orthogonal-relation} we first note the
basis components ${v}_j^{(m\,h\,k)}$ in global coordinates have the
$\tau$ and $\varphi$ dependence,
\begin{equation}
{v}_j^{(m\,h\,k)} \sim \exp{(i m\varphi)\exp{\left[i(h-k)\tau\right]}}.
\label{eq:vector-basis-general-form}
\end{equation}
This dependence on $\tau$ and $\varphi$ is the same for the scalar and
tensor basis components.  Once we integrate over $\varphi$ and $\tau$
in Eq.~\eqref{eq:orthogonal-relation}, the integral will be
proportional to
$\delta_{m,m^\prime}\delta_{h-k,h^\prime-k^\prime}$. Notice that the
boundaries $\tau\rightarrow \pm \infty$ are oscillatory, so the $\tau$
integral needs to be regulated in the same way as Fourier integrals.

Now we only need to show bases with different weight $k$ are
orthogonal.  Once this is done we will recover
Eq.~\eqref{eq:orthogonal-relation}.  For simplicity, from now on we
only track the $k$-index in the vector bases.  Recall that we obtain
the lower weight bases by applying the lowering operator order by
order,
\begin{align}
\langle {\bf u}^{(k)}, {\bf v}^{(k^\prime)} \rangle
= \langle {\bf u}^{(k)}, \mathcal{L}_{L_-}{\bf v}^{(k^\prime-1)}
  \rangle
\,.
\end{align}
Now we try to ``integrate by parts'' with the Lie derivative,
\begin{widetext}
\begin{align}\label{eq:boundary-extraction}
\langle {\bf u}^{(k)}, \mathcal{L}_{L_-}{\bf v}^{(k^\prime-1)}
  \rangle
&= \int_{\Sigma_u} \mathcal{L}_{L_-}\left(\overline{u_i^{(k)}} v^i_{(k^\prime)}\right) \dvol- \langle \overline{\mathcal{L}_{L_-}}{\bf u}^{(k)}, {\bf v}^{(k^\prime-1)} \rangle, \\
&= \int_{\Sigma_u} \mathcal{L}_{L_-}\left(\overline{u_i^{(k)}} v^i_{(k^\prime)}\right) \dvol + \langle \mathcal{L}_{L_+}{\bf u}^{(k)}, {\bf v}^{(k^\prime-1) } \rangle,
\end{align}
where in the last line we used the fact that $\overline{L_+}=-L_-$.
Note that this relationship does not hold between $H_{\pm}$, so this
type of proof will not work in Poincar\'e coordinates.

We would like to discard the first term on the RHS of
Eq.~\eqref{eq:boundary-extraction}, which would show that
$\lie_{L_+}$ and $\lie_{L_{-}}$ are adjoints of each other.  We can
do this by converting the Lie derivative into a covariant derivative
and then a total divergence.  Since $L_{\pm}$ are KVFs, they are
automatically divergence-free, so we can pull them inside the
covariant derivative:
\begin{equation}
\int_{\Sigma_u} \dvol \mathcal{L}_{L_-}\left(\overline{u_i^{(k)}} v^i_{(k^\prime)}\right)
= 
\int_{\Sigma_u} \dvol L^j_- D_j \left( \overline{u_i^{(k)}} v^i_{(k^\prime)}\right)
= 
\int_{\Sigma_u} \dvol D_j \left( L^j_-\overline{u_i^{(k)}} v^i_{(k^\prime)}\right)
\,.
\end{equation}
\end{widetext}
This step is identical if we are considering scalars/vectors/tensors,
since the argument of the Lie derivative has all indices contracted.
Using Stokes' theorem, the integral of the total derivative becomes a
boundary integral, evaluated at $\psi=0,\pi$.  This boundary
contribution vanishes for $h<-1$ in the highest-weight module.
To see this, one must count the powers of $\sin\psi$ which depends on
$h$ (see App.~\ref{app:global-basis}), and take into account the
volume element's contribution, $\sqrt{-\gamma}\propto(\sin\psi)^{-2}$.

We repeat the procedure of extracting lowering operators from the ket
as in Eq.~\eqref{eq:boundary-extraction}, and arrive at
\begin{equation}
\langle {\bf u}^{(k)}, {\bf v}^{(k^\prime)} \rangle = \langle
\left(\mathcal{L}_{L_+}\right)^{k^\prime}{\bf u}^{(k)}, {\bf v}^{(0)} \rangle
\,.
\end{equation}
Recall that the vector basis terminates at the highest weight.
Therefore when $k^\prime > k$, $\left(\mathcal{L}_{L_+}\right)^{k^\prime} {\bf u}^{(k)}$ will
vanish.  Similarly when $k^\prime < k$, we can extract all lowering
operators from the bra and raise the weight of the states in the ket,
which will terminate upon raising the highest-weight state. Therefore
the vector bases with different weights $k, k^\prime$ are
orthogonal.

Since we have also proved that vector bases with different $m$ and
$h-k$ are orthogonal, the proof of orthogonality for vector bases is
done.  It may not be obvious that the proof holds unaltered for
scalars/vectors/tensors.  In all the relevant steps above, we have
noted where each argument works for each of the three types of fields.

Therefore we arrive at the conclusion that the scalar, vector, and
symmetric tensor bases in global coordinates form orthogonal basis
sets.\hfill{}$\qed$

\section{Separation of variables}
\label{sec:separation-variables}

In this section we show that with the scalar, vector, and tensor bases
we have obtained, it is possible to separate variables for many
physical systems in NHEK spacetime.  One can show that all conclusions
in this section hold for both Poincar\'{e} coordinates and global
coordinates.  In global coordinates the results are in general more
complicated.  Therefore for concreteness all results in this section
are given in Poincar\'{e} coordinates.

The main result of this section can be summarized with the schematic
equation:
\begin{align}
  \mathcal{D}_{x}
  \left[
  \left(
  \parbox{2.cm}{\centering\tiny$\SLTRU$\\ structure ($T, \Phi, R$)}
  \right)^{(m,h,k)}
  \times
  \left(
  \parbox{1.5cm}{\centering\tiny$u$ (or $\cos\theta$)\\ dependence}
  \right)
  \right]
  = \nn \\
  \left(
  \parbox{2.cm}{\centering\tiny$\SLTRU$\\ structure ($T, \Phi, R$)}
  \right)^{(m,h,k)}
  \times
  \mathcal{D}^{(m,h)}_{u}
  \left[
  \parbox{1.5cm}{\centering\tiny$u$ (or $\cos\theta$)\\ dependence}
  \right]
  \,. \nn
\end{align}
Here, $\mathcal{D}_{x}$ is an $\SLTRU$-equivariant differential
operator, which takes derivatives in the $T,\Phi,R,u$ directions.  We
completely specify the $T,\Phi,R$ dependence by being in a certain
irreducible representation (irrep)
of $\SLTRU$ labeled by $(m,h,k)$.  Then the $\SLTRU$ structure
factors straight through the differential operator $\mathcal{D}_{x}$,
leaving a new differential operator $\mathcal{D}_{u}^{(m,h)}$ which
only takes $u$ derivatives.  This greatly simplifies computations,
since the partial differential equations have been converted into
ordinary differential equations (ODEs).  Because of the
$\SLTRU$-invariance, notice that $\mathcal{D}_{u}^{(m,h)}$ only
depends on $m$ and $h$, which label the irrep, and not on $k$, which
labels the descendant number within the irrep.

\subsection{Covariant differentiation preserves\break{} isometry group irrep labels}
\label{sec:general-statement}

Let us first make a general statement about how the presence of a
group of isometries acting on the manifold can be useful in separation
of variables.  The conclusions obtained in this subsection will also
justify our motivations of finding group representations for NHEK's
isometry.  Consider a manifold $\mathcal{M}$ with metric $g_{ab}$,
metric-compatible connection $\cd$, and an isometry Lie group $G$
acting on the manifold.  Let $\alpha_{(i)}\in \mathfrak{g}$ be a basis for
the Lie algebra, with representation $\{X_{(i)}\}$ on the manifold.
Further, let $c^{(i)(j)}$ be the inverse of the Killing form of the
Lie algebra in this basis~\cite{Barut:1986dd}.  Then we also have a
quadratic Casimir element, which acts on any tensor $\mathbf{t}$ as
\begin{align}
  \Omega \cdot \mathbf{t} \equiv \sum_{i,j} c^{(i)(j)}
  \lie_{X_{(i)}} \lie_{X_{(j)}} \mathbf{t}
  \,.
\end{align}
Irreps of $G$ will be labeled by eigenvalues $\lambda_{i}$
of \emph{some} of the KVFs, and the eigenvalue $\omega$ of the Casimir
$\Omega$.

First, we need a lemma on the commutation relation of manifold
isometries and covariant derivatives,
\begin{equation}
\left[ \mathcal{L}_{X_{(i)}} , \nabla_a \right]{\bf t}=0,
\label{eq:commutation-isometry-covariant}
\end{equation} 
where $\bf t$ can be a scalar, vector, or tensor.  To prove
Eq.~\eqref{eq:commutation-isometry-covariant}, one can start by
showing the commutation relations for $\bf t$ being a 0-form (which
follows immediately from Cartan's magic formula for a 0-form) and a
one-form, then use the Leibniz rule to generalize the relations to the
vector and tensor cases.
Eq.~\eqref{eq:commutation-isometry-covariant} says that the operator
$\cd_{a}$ is $\SLTRU$ \emph{equivariant}: that is, its action commutes
with left-translation by the group~\cite{MR2954043}.

An important consequence of the commutation relation
Eq.~\eqref{eq:commutation-isometry-covariant} is that the Casimir
element $\Omega$ of the algebra $\mathfrak{g}$ also commutes with the covariant
derivative.  Simply commute each Lie derivative one at a time, and
the coefficients $c^{(i)(j)}$ are constants.  As a result,
\begin{equation}
\left[ \Omega , \nabla_a \right]\mathbf{t}=0.
\label{eq:commutation-casimir-covariant}
\end{equation} 

Now consider a tensor $\mathbf{t}$ living in an irrep with labels $\lambda_{i}$
and $\omega$, meaning
\begin{align}
  \lie_{X_{(i)}} \mathbf{t} &= \lambda_{i} \mathbf{t} \,, \\
  \Omega \cdot \mathbf{t} &= \omega \mathbf{t} \,.
\end{align}
As an immediate consequence of
Eq.~\eqref{eq:commutation-isometry-covariant} and
Eq.~\eqref{eq:commutation-casimir-covariant} is that $\cd \mathbf{t}$
has the same labels $\lambda_{i}$ and $\omega$,
\begin{align}
 \lie_{X_{(i)}} \cd\mathbf{t} &= \lambda_{i} \cd\mathbf{t} \,, \\
 \Omega \cdot \cd \mathbf{t} &= \omega \cd\mathbf{t} \,.
\end{align}

Thus any linear differential operator which is built just from $\cd_{a}$ and
the metric $g_{ab}$ can not mix tensors with different irrep labels
$(\lambda_{i}, \omega)$.  This even extends to differential operators
which include the Levi-Civita tensor $\epsilon$ and the Riemann tensor
$R_{abcd}$, because these two objects are also annihilated by all of
the $\lie_{X_{(i)}}$.  As a result, when tensors are decomposed into a
sum over irreps with different labels, they will remain separated in
the same ways under this type of differential operator.  This is the
underlying reason why the method of finding the unitary irreps of
NHEK's isometry introduced in Sec.~\ref{sec:high-lowest-weight} will
lead to separation of variables in many physical systems.

\subsection{Scalar Laplacian}
\label{sec:sep-scalar}

As the first example, we look at the massless scalar wave equation
$\Box \psi = S$ in NHEK space time, where $S$ is a source term
(including a mass term also works).
Since the scalar d'Alembert operator $\Box \equiv \cd^{a}\cd_{a}$ is
built only from $g_{ab}$ and $\cd_{a}$, it should commute with
$\Omega$ and $\lie_{X}$ where $X$ is any KVF.  To show this
explicitly, note that in Poincar\'e coordinates, $\Box\psi$ can be
written as
\begin{align}
\Box \psi={}&\frac{1}{2\Gamma(u)}\Bigg\{(\Omega
        + \Xi(u)\lie^2_{Q_0})\psi +\lie_{\partial _u}\left[(1-u^2)\lie_{\partial _u}\psi\right]\Bigg\}
  \,,
\end{align}
where $\Xi(u)\equiv\Lambda(u)^{-2}-1$.

Assume we can decompose an arbitrary scalar field
$\psi(T,\,\Phi,\,R,\,u)$ according to
\begin{align}
  \psi &= \sum_{mhk}C_{mhk}(u)F^{(m\,h\,k)}(T,\,\Phi,\,R) \\ \nonumber
       &= \sum_{mhk} \psi_{mhk}(T,\,\Phi,\,R,\,u), 
\end{align}
where $F$ is the scalar basis on $\Sigma_u$ and $C_{mhk}$ are some
unknown functions of $u$.  We also decompose the source term using the
scalar basis functions via $S=\sum_{mhk} S_{mhk} F^{(m\,h\,k)}$.  The basis
functions $F^{(m\,h\,k)}$ are eigenfunctions of $\Omega$ and
$\lie_{Q_{0}}$, and so $\psi_{mhk}$ are also eigenfunctions.
Therefore it is straightforward to see that the
$(T,\Phi,R)$-dependence in $\psi_{mhk}$ is invariant after applying
the scalar box operator.  The equation for a specific mode labeled by
$(m,h,k)$ becomes
\begin{align}
 &S_{mhk} F^{(m\,h\,k)}={}\Box^{(m,h)} \psi_{mhk}={}\frac{1}{2\Gamma(u)}\times \\\nn
  &\times\Bigg\{[h(h+1)
  - m^{2} \Xi(u)]\psi_{mhk} +\mathcal{L}_{\partial _u}\left[(1-u^2)\mathcal{L}_{\partial _u}\psi_{mhk}\right]\Bigg\} 
\,.
\end{align}
This entire equation is proportional to the basis function
$F^{(m\,h\,k)}$, which can thus be divided out, leaving an ODE for one
function, $C_{mhk}(u)$.

Specializing to the homogeneous (source-free) case, we find the ODE
\begin{align}
  \frac{d}{du}
  \left[
  (1-u^{2}) \frac{d}{du}
  C_{mhk}
  \right]
  + \left[h(h+1) - \Xi(u) m^2 \right]C_{mhk}=0
\,.
\end{align}
This equation has two regular singularities $u=\pm 1$ and an irregular
singularity of rank 1 at $u=\infty$, which falls into the class of
confluent forms of Heun's equation~\cite{NIST:DLMF}.  Explicitly, it
is a spheroidal differential equation, whose standard form is
\begin{equation}
  \frac{\text{d}}{\text{d}u}\left( (1-u^2)\frac{\text{d}\varphi}{\text{d}u}\right) 
  + \left(\lambda + \gamma^2(1-u^2)-\frac{\mu^2}{1-u^2}\right)\varphi = 0,
  \label{eq:spheroidal-equation}
\end{equation}
where we have made the substitution $\lambda = h(h+1)+2m^2$,
$\gamma^2 = -m^2/4$ and $\mu^2 = m^2$.  When $\gamma = 0$,
Eq.~\eqref{eq:spheroidal-equation} reduces to the Legendre
differential equation and the solutions are Legendre polynomials.
Being second order, the space of solutions is two dimensional,
\begin{equation}
  \varphi(u) = a_1S^{(1)}_{n\mu}(\gamma,u) + b_1S^{(2)}_{n\mu}(\gamma,u) .
\end{equation}
A solution that is regular at $u=\pm 1$ only exists for
eigenvalues $\lambda = \lambda_n^m(\gamma^2)$, where
$\mu = m = 0, 1, 2, \ldots,$ and $n = m, m+1, m+2,\ldots$.  Thus,
there are only discrete values of the irrep label $h$ which satisfy
regularity at the poles $u=\pm 1$.

\subsection{Maxwell system}
\label{sec:sep-vector}

Let's look at another system of physical importance, the Maxwell
system, and verify that we can separate variables in Maxwell's
equations (the Proca equation---i.e.~adding a mass term---works as
well).  The inhomogeneous Maxwell equations in the presence of a
source vector field $\mathcal{J}$ are
\begin{equation}
\nabla^{a}\mathcal{F}_{ab} = \mathcal{J}_{b},
\end{equation}
where the electromagnetic tensor $\mathcal{F}$ is built from the vector potential $A$ according to 
\begin{equation}
\mathcal{F}_{ab} = \nabla_a A_b - \nabla_b A_a.
\end{equation}
We again assume that we can expand the vector potential in the scalar
and vector bases.  Define a one-form $n_a=\text{d}u$, this
expansion is given by
\begin{equation}
  \label{eq:vector-potential}
  A_{a} = \sum_{mhk}\left(C_u(u) n_{a} F^{(m\,h\,k)}
    + \sum_{B} C_B(u){V_{a}^{B}}^{(m\,h\,k)}\right),
\end{equation}
where $B\in \{T,\Phi,R\}$, $C_B(u)$ and $C_u(u)$ are unknown functions
of $u$.  Notice that $B$ is \emph{not} a tensor index.  It is the
label of a specific choice of vector bases and their corresponding
unknown $C$-functions.  The expression of $F^{(m\,h\,k)}$ and the
projection of ${V_{a}^{B}}^{(m\,h\,k)}$ onto $\Sigma_u$,
i.e.~${V_{i}^{B}}^{(m\,h\,k)}$ are both given in App.~\ref{app:Poincare-basis}.
Then at the highest weight $k=0$, the left hand side of Maxwell's equation can be
rewritten as
\begin{align}
  \nabla^{a}\mathcal{F}_{ab}|_{k=0}
  &= \mathcal{D}^{(m,h)}_u[\mathbf{C}(u)] n_{b} F^{(m\,h\,0)} \\ \nonumber
  &+ \sum_{B}
    \mathcal{D}^{(m,h)}_{B}[\mathbf{C} (u)]V_b^{B(m\,h\,0)}
    \,,
\end{align}
where we have collected the four $C$-functions into the vector
$\mathbf{C}(u)$, and defined the general differentiation as
$\mathcal{D}^{(m,h)}[\mathbf{C} (u)]$, whose expressions are given in
App.~\ref{app:G-function}.   As long as the source field can also be
decomposed using the scalar and vector bases, the inhomogeneous
Maxwell equations in NHEK will reduce to four ordinary differential
equations with four unknown $C$-functions.  Although we only show this
is true for the highest-weight case, this conclusion holds for any
$k$.  This is due to the commutation of the lowering operator and the
covariant differentiation.  For explicit calculations of Maxwell's
system using the highest-weight vector basis we refer our readers
to~\cite{Lupsasca:2014pfa, Compere:2015pja}.

\subsection{Linearized Einstein system}
\label{sec:sep-symmetric-tensor}

In this subsection we show that we can separate variables on the left
hand side of linearized Einstein equation, using our scalar, vector,
and tensor bases for NHEK.  Consider the metric perturbation
$g^\prime_{ab} = g_{ab} + \epsilon h_{ab}+\mathcal{O}(\epsilon^{2})$,
where $g_{ab}$ is the NHEK metric
and $h_{ab}$ is a perturbation.  The linearized Einstein equations
(i.e.~at order $\epsilon^{1}$) are
\begin{equation}
  G^{(1)}_{ab}[ h ] =  8 \pi T_{ab}
\,,
\end{equation}
where $T_{ab}$ is the stress-energy tensor of a source term.  The
linearized Einstein operator $G^{(1)}[h]$ can be written in terms of
the background covariant derivative $\cd$ as
\begin{align}
-2 G^{(1)}_{ab}[h] ={}&
\Box \overline{h}_{ab}
+ g_{ab} \cd^{c}\cd^{d}\overline{h}_{cd}
-2\cd^{c}\cd_{(a}\overline{h}_{b)c}
\nn \\ &
- g_{ab} R^{cd}\ \overline{h}_{cd} +
R\ \overline{h}_{ab}
\,,
\end{align}
where $\overline{h}_{ab} = h_{ab}-\frac{1}{2}g_{ab}g^{cd}h_{cd}$ is
the trace-reverse of $h_{ab}$, $R_{ab}$ is the background Ricci
curvature, $R$ is the background Ricci scalar, and parentheses around $n$ indices
means symmetrizing with a factor of $1/n!$.  This operator, again, is
$\SLTRU$ equivariant.

We assume that we can expand the metric perturbation in our
scalar, vector, and tensor bases, according to
\begin{align}
  h_{ab} &= \sum_{mhk} h^{(m\,h\,k)}_{ab}
         =\sum_{mhk}\Bigg(n_a n_b F^{(m\,h\,k)}C_{uu}(u)   \\ \nonumber
         &+ \sum_{B} 2n_{(a\vphantom{b)}}^{\vphantom{A(}}V_{b)}^{B(m\,h\,k)} C_{uB}(u) 
         + \sum_{A,B} W_{ab}^{AB(m\,h\,k)} C_{AB}(u) \Bigg),
\end{align}
where $A,B\in\{T,\Phi,R\}$, $C_{uu},C_{uB},C_{AB}$ are unknown
functions of $u$. 
Notice that $A$ and $B$ are \emph{not} tensor indices
but only labels of a specific choice of the vector and tensor bases (introduced in 
App.~\ref{app:vector-basis-Poincare} and~\ref{app:tensor-basis-Poincare}) and their corresponding unknown $C$-functions. Thus there are 
no differences between a subscript and a superscript $A$ or $B$.  We choose the three highest-weight vector bases
$V_b^{B(m\,h\,0)}$ and the six highest-weight tensor bases $W_{ab}^{AB(m\,h\,0)}$ such that the
metric perturbation with $k=0$ can be written as
Eq.~\eqref{eq:hab-highest}.  We substitute the
highest-weight metric perturbation into the left hand side of
the linearized Einstein equation and the result is given by
Eq.~\eqref{eq:G1-hab}.
\begin{widetext}
\begin{align}{}
h^{(m\,h\,0)}_{ab} &= 
  R^h e^{im\Phi}
  \left[
  \begin{array}{cccc}
   R^{+2}C_{TT}(u)    
   & R^{+1} C_{T\Phi}(u)   
   & R^{+0}C _{TR}(u)       
   & R^{+1} C_{uT}(u) \\
 * & R^{+0} C_{\Phi\Phi}(u) 
   & R^{-1}C_{R\Phi}(u)
   & R^{+0} C_{u\Phi}(u)  \\
 * &                                     
 * & R^{-2}C_{RR}(u)     
   & R^{-1}C_{uR}(u)  \\
 * &                                     
 * &                               
 * & R^{+0} C_{uu}(u)  
 \end{array}
 \right]
 \label{eq:hab-highest}
\\
\intertext{}
G^{(1)}_{ab} [ h^{(m\,h\,0)} ] &=
  R^h e^{im\Phi}
  \left[
  \begin{array}{cccc}
   R^{+2}\mathcal{D}^{(m,h)}_{TT}[\mathbf{C}(u)]    
   & R^{+1}\mathcal{D}^{(m,h)}_{T\Phi}[\mathbf{C}(u)]   
   & R^{+0}\mathcal{D}^{(m,h)}_{TR}[\mathbf{C}(u)]        
   & R^{+1}\mathcal{D}^{(m,h)}_{uT}[\mathbf{C}(u)]  \\
 * & R^{+0}\mathcal{D}^{(m,h)}_{\Phi\Phi}[\mathbf{C}(u)]  
   & R^{-1}\mathcal{D}^{(m,h)}_{R\Phi}[\mathbf{C}(u)]  
   & R^{+0}\mathcal{D}^{(m,h)}_{u\Phi}[\mathbf{C}(u)]  \\
 * &                                     
 * & R^{-2}\mathcal{D}^{(m,h)}_{RR}[\mathbf{C}(u)]     
   & R^{-1}\mathcal{D}^{(m,h)}_{uR}[\mathbf{C}(u)]  \\
 * &                                     
 * &                               
 * & R^{+0}\mathcal{D}^{(m,h)}_{uu}[\mathbf{C}(u)]  
 \end{array}
 \right]
 \label{eq:G1-hab}
\end{align}
\end{widetext}
Again notice that the $(T,\Phi,R)$ dependence has factored straight
through the differential operator, resulting in ten coupled ODEs for
the ten $C$-functions, which we have collected together as
$\mathbf{C}(u)$.  The expressions for all these differential operators
are given in App.~\ref{app:A-B-C-matrice}.

We can easily verify that $G^{(1)}$ commutes with $\mathcal{L}_{H_-}$,
therefore the linearized Einstein operator acting on a basis function
with arbitrary weight can be obtained easily by repeatedly applying
the lowering operator $\mathcal{L}_{H_-}$, $k$ times, on
Eq.~\eqref{eq:G1-hab}.  While applying the lowering operator, in
general different components of $G^{(1)}_{ab}[h^{(m\,h\,k)}]$ will get
mixed up, but the separation of variables still holds.  Therefore we
conclude that with these scalar, vector, and tensor bases, we can
separate variables in the linearized Einstein system in NHEK.

Given some source terms, these bases can be directly applied to
solving for the corresponding metric perturbations.  For instance, we
have obtained the highest-weight metric deformations in NHEK sourced
by the decoupling limits of dynamical Chern-Simons and
Einstein-dilaton-Gauss-Bonnet gravity~\cite{ChenSteinForthcoming}.

\section{Conclusions and future work}
\label{sec:concl-future-work}

In this paper, we proposed an isometry-inspired method to study metric
perturbations in the near-horizon extremal Kerr spacetime.  That is,
we separated variables in the metric perturbation equations in the
NHEK spacetime, by expanding the perturbation in terms of basis
functions adapted to the isometry group.  With the separable
linearized Einstein equation, one obtains the perturbed metric
directly, without the complication of metric reconstruction.  Further,
our formalism does not depend on gauge choice.  Within our formalism,
partial differential equations built from $\SLTRU$-equivariant operators
can be converted into ordinary differential equations in the polar
angle, which are simpler to solve.
The price is that one must solve coupled, rather than decoupled,
equations in our metric formalism.

We accomplished three things: (i) we used the highest-weight method to
obtain the scalar, vector, and symmetric tensor bases for the isometry
group of NHEK; (ii)~in global coordinates, we showed that these bases
form orthogonal basis sets when the labels of irreps satisfy $h<-1$;
and (iii)~with these basis functions, we separated variables in many
physical equations like the scalar wave equation, Maxwell's equations,
and the linearized Einstein equations.

\paragraph*{Future work.}
Although we have shown that bases in global coordinates are
orthogonal, we did not mention completeness.  There are clues that, in
global coordinates, combining the highest- and lowest-weight modules
will give a complete set of states.  We leave a rigorous treatment of
completeness to future work.  However, many problems can already be
attacked without worrying about completeness---for example, if the
source term lives in exactly one irrep.

Since the near-horizon near-extremal geometry exhibits the same
isometry as NHEK, we expect all discussions in this paper can be
applied to understanding metric perturbations in near-NHEK, which is
more astrophysically relevant.  With the knowledge of isometry-adapted
bases in NHEK, we hope to enhance our understanding of the Kerr/CFT
conjecture~\cite{Guica:2008mu} from the gravity side.

\acknowledgments
The authors would like to thank
Yanbei Chen,
Alex Lupsasca,
Zachary Mark,
and
Peter Zimmerman
for useful conversations.
LCS acknowledges the support of NSF grant PHY--1404569, and both
authors acknowledge the support of the Brinson Foundation.  Some 
calculations used the computer algebra system \textsc{Mathematica},
in combination with the \textsc{xAct/xTensor}
suite~\cite{JMM:xAct,MARTINGARCIA2008597}.

\begin{appendix}
\section{Scalar, vector, and\break{} symmetric tensor bases}
\label{app:S-V-T-Basis}

In this section we present the expressions of scalar, vector, and
symmetric tensor bases both in Poincar\'{e} coordinates and global
coordinates, up to constant factors.  All the basis functions are
defined on the three-dimensional hypersurface $\Sigma_u$.  To promote
these basis functions to the full four-dimensional manifold
$\mathcal{M}$, one promotes all constant coefficients $c_{\beta}$ to become
unknown functions of the (cosine) polar angle, $c_{\beta}(u)$.  The basis
functions given here are (mostly) obtained using the highest-weight
method introduced in Sec.~\ref{sec:high-lowest-weight}, i.e.~they form
the highest-weight modules for $\SLTRU \circlearrowleft \mathcal{M} $.
Such a highest-weight module is infinite dimensional, the length of
this paper, however, is supposed to be finite.  Therefore, we give the
highest three weights for scalar bases, the highest two weights for
vector bases, and only the highest weight for tensor bases.  Note all
other basis functions can be generated by applying the lowering
operator on the highest weight basis order by order.  In order to
compare the basis functions in different modules, in global
coordinates, we also give the expressions of the scalar bases obtained
using the lowest-weight method.

All expressions in these appendices are also available in the
companion \textsc{Mathematica} notebooks:
\texttt{Sep-met-pert-in-NHEK-Poinc.nb},
\texttt{Sep-met-pert-in-NHEK-global.nb},
and precomputed quantities in
\texttt{NHEK-precomputed.mx} \cite{NHEKsupplement}.

\subsection{Basis functions in Poincar\'{e} coordinates}
\label{app:Poincare-basis}

\subsubsection{Scalar bases}
\label{app:scalar-basis-Poincare}

The scalar bases in Poincar\'{e} coordinates are given by 
\begin{equation}
F^{(m\,h\,k)} \propto  R^{h-k} e^{i m \Phi} \times f^{(m\,h\,k)}\,,
\end{equation}
where 
\begin{align}
f^{(m\,h\,0)} =& 1\,, \\ \nn
f^{(m\,h\,1)} =& -2 (h R T+i m)\,, \\ \nn
f^{(m\,h\,2)} =& -2 [-2 i (2 h-1) m R T + \\\nn
&\quad{}+h (1-2 h) R^2 T^2+h+2 m^2]\,.
\end{align}

\subsubsection{Vector bases}
\label{app:vector-basis-Poincare}

The covector bases in Poincar\'{e} coordinates can be decomposed
using the dual basis one-forms $\{\text{d}T,\text{d}\Phi,\text{d}R\}$
via
\begin{equation}
\mathbf{V}^{(m\,h\,k)}= V_{i}^{(m\,h\,k)}\text{d}x^i,\quad x\in\{T,\Phi,R\}\,.
\end{equation}
The covector components are given by
\begin{align}
  V_{i}^{(m\,h\,k)} \propto
  \begin{bmatrix}
    v^{(m\,h\,k)}_{T}R^{+1} \\
    v^{(m\,h\,k)}_{\Phi} R^{+0} \\
    v^{(m\,h\,k)}_{R} R^{-1}
  \end{bmatrix}
   R^{h-k} e^{i m \Phi}
   \,,
\end{align}
where
\begin{align}
v_{T}^{(m\,h\,0)} &= c_1\,, &
v_{\Phi}^{(m\,h\,0)} &= c_2\,, &
v_{R}^{(m\,h\,0)} &= c_3\,, &
\end{align}
and 
\begin{align}
v_{T}^{(m\,h\,1)} &= -2 [c_3+c_1 (h R T+i m)]\,, \\ \nn
v_{\Phi}^{(m\,h\,1)} &= -2 c_2 (h R T+i m)\,, \\ \nn
v_{R}^{(m\,h\,1)} &= -2  [c_3 (h R T+i m)+c_1-c_2]\,.
\end{align}
Notice that there are three unknown coefficients $c_1$, $c_2$, and
$c_3$.  They endow us the freedom of choosing a 3-dimensional basis
for covectors.
In particular, we introduce a specific set of covector bases labeled by
$B$ where $B\in\{T,\Phi,R\}$. They are defined by
\begin{align}
\mathbf{V}_T^{(m\,h\,k)}&=\mathbf{V}^{(m\,h\,k)}|_{c_2=c_3=0}\,, \\ \nn
\mathbf{V}_{\Phi}^{(m\,h\,k)}&=\mathbf{V}^{(m\,h\,k)}|_{c_1=c_3=0}\,,  \\ \nn
\mathbf{V}_R^{(m\,h\,k)}&=\mathbf{V}^{(m\,h\,k)}|_{c_1=c_2=0}\,.
\label{eq:specific-choice-of-vector}
\end{align}

\subsubsection{Symmetric tensor bases}
\label{app:tensor-basis-Poincare}

The symmetric tensor bases in Poincar\'{e} coordinates can be decomposed using the dual basis one-forms $\{\text{d}T,\text{d}\Phi,\text{d}R\}$ via
\begin{equation}
\mathbf{W}^{(m\,h\,k)} = W^{(m\,h\,k)}_{ij}\,\text{d}x^i\otimes \text{d}x^j,\quad x\in\{T,\Phi,R\}\,.
\end{equation}
The tensor components are given by
\begin{align}
W_{ij}^{(m\,h\,k)} \propto 
  \left[
  \begin{array}{ccc}
   R^{+2}w^{(m\,h\,k)}_{TT}
   & R^{+1} w^{(m\,h\,k)}_{T\Phi}
   & R^{+0} w^{(m\,h\,k)}_{TR}      \\
 * & R^{+0} w^{(m\,h\,k)}_{\Phi\Phi}
   & R^{-1} w^{(m\,h\,k)}_{R\Phi}     \\
 * &
 * & R^{-2} w^{(m\,h\,k)}_{RR}
 \end{array}
 \right] \times \nn \\
  \times R^{h-k} e^{im\Phi}\,,
 \end{align}
where 
\begin{align}
w_{TT}^{(m\,h\,0)} &=c_1\,, &
w_{\Phi\Phi}^{(m\,h\,0)} &= c_2\,,&
w_{RR}^{(m\,h\,0)} &= c_3\,,\\ \nn
w_{T\Phi}^{(m\,h\,0)} &= c_4\,, &
w_{\Phi R}^{(m\,h\,0)} &= c_5\,,&
w_{RT}^{(m\,h\,0)} &= c_6\,.
\end{align}
Notice that there are six unknown $c$-coefficients. They endow us 
the freedom of choosing the six tensor bases. In particular, we 
introduce a specific set of highest-weight tensor bases labeled by
$A,B$ where $A,B\in\{T,\Phi,R\}$. They are defined by
\begin{align}
\mathbf{W}_{TT}^{(m\,h\,k)}&=\mathbf{W}^{(m\,h\,k)}\big|_{c_{\beta\neq 1}=0}\,, \\ \nn
\mathbf{W}_{\Phi\Phi}^{(m\,h\,k)}&=\mathbf{W}^{(m\,h\,k)}\big|_{c_{\beta\neq 2}=0}\,,  \\ \nn
\mathbf{W}_{RR}^{(m\,h\,k)}&=\mathbf{W}^{(m\,h\,k)}\big|_{c_{\beta\neq 3}=0}\,, \\ \nn
\mathbf{W}_{T\Phi}^{(m\,h\,k)}&=\mathbf{W}^{(m\,h\,k)}\big|_{c_{\beta\neq 4}=0}\,, \\ \nn
\mathbf{W}_{\Phi R}^{(m\,h\,k)}&=\mathbf{W}^{(m\,h\,k)}\big|_{c_{\beta\neq 5}=0}\,,  \\ \nn
\mathbf{W}_{RT}^{(m\,h\,k)}&=\mathbf{W}^{(m\,h\,k)}\big|_{c_{\beta\neq 6}=0}\,.
\label{eq:specific-choice-of-tensor}
\end{align}
This specific choice of tensor bases will be utilized to write the metric perturbation as 
in Eq.~\eqref{eq:hab-highest}.

\begin{widetext}

\subsection{Basis functions in global coordinates}
\label{app:global-basis}

\subsubsection{Scalar bases (highest-weight module)}
\label{app:scalar-basis-highest-reps}
The scalar bases from the highest-weight module in global coordinates are given by
\begin{equation}
F^{(m\,h\,k)} \propto  (\sin\psi)^{-h}  e^{i [(h-k) \tau + m \varphi] +m \psi} \times f^{(m\,h\,k)}\,,
\end{equation}
where
\begin{align}
f^{(m\,h\,0)} &=1 \,, \\ \nn
f^{(m\,h\,1)} &=-2 (m \sin \psi -h \cos \psi ) \,, \\ \nn
f^{(m\,h\,2)} &= 2 \left[h^2+m^2 +\left(h^2-h-m^2\right) \cos 2 \psi +(m-2 h m) \sin 2 \psi \right]\,.
\end{align}

\subsubsection{Scalar bases (lowest-weight module)}
\label{app:scalar-basis-lowest-reps}
The scalar bases from the lowest-weight module in global coordinates are given by
\begin{equation}
F_L^{(m\,h\,k)} \propto  (\sin\psi)^{+h}  e^{i [(h+k) \tau + m \varphi] - m \psi} \times f_L^{(m\,h\,k)}\,,
\end{equation}
where 
\begin{align}
f_L^{(m\,h\,0)} &=1 \,, \\ \nonumber
f_L^{(m\,h\,1)} &= -2 (m \sin \psi -h \cos \psi ), \\ \nonumber
f_L^{(m\,h\,2)} &= 2  \left[h^2+m^2 +\left(h^2+h-m^2\right) \cos 2 \psi -(m+2 h m) \sin 2 \psi \right].
\end{align}

\subsubsection{Vector bases}
\label{app:vector-basis-global}

The covector bases in global coordinates can be decomposed using the
dual basis one-forms $\{\text{d}\tau,\text{d}\varphi,\text{d}\psi\}$
via
\begin{equation}
\mathbf{V}^{(m\,h\,k)}= V_{i}^{(m\,h\,k)}\text{d}x^i,\quad x\in\{\tau,\varphi,\psi\}\,.
\end{equation}
The covector components are given by
\begin{align}
  V_{j}^{(m\,h\,k)} \propto
  \begin{bmatrix}
    v^{(m\,h\,k)}_{\tau}(\sin \psi )^{-1} \\
    v^{(m\,h\,k)}_{\varphi} (\sin \psi )^{+0} \\
    v^{(m\,h\,k)}_{\psi} (\sin \psi )^{-1}
  \end{bmatrix}
  (\sin \psi )^{-h} e^{i [(h-k) \tau + m \varphi] +m \psi}
  \,,
\end{align}
where
\begin{align}
v_{\tau}^{(m\,h\,0)} &=   -\frac{1}{4 }\left(c_1 e^{- i \psi }+ 2c_1 e^{i \psi }-2 c_2 e^{ -i \psi } +4 c_3 e^{ i \psi }\right)\,, \\ \nn
v_{\varphi}^{(m\,h\,0)} &= c_1 \,, \\ \nn
v_{\psi}^{(m\,h\,0)} &= +\frac{1}{4 }\left(c_1 e^{- i \psi }+2 c_2 e^{ -i \psi } +4 c_3 e^{ i \psi }\right) \,,
\end{align}
and 
\begin{align}
v_{\tau}^{(m\,h\,1)} =& -\frac{1}{4}\bigg\{
c_1 [2 (h+i m)e^{2 i \psi } +(3 h-i m-1)+(h-i m+1)e^{- 2i \psi }]-\\ \nn
&-2c_2 [ (h+i m+1)+ (h-i m-1)e^{-2 i \psi }]
+4 c_3 [(h+i m-1)e^{2i \psi } +(h-i m+1)]
\bigg\}\,,\\  \nn
v_{\varphi}^{(m\,h\,1)} =&  -2 c_1(m \sin \psi -h \cos \psi )\,, \\ \nn
v_{\psi}^{(m\,h\,1)} =& +\frac{1}{4} \bigg\{
c_1 [(h+i m+1)+(h-i m-1)e^{-2 i \psi } ]+ 
2 c_2 [(h+i m+1)+(h-i m-1)e^{-2 i \psi } ]\\\nn
&+4 c_3 [(h+i m-1)e^{2 i \psi} +(h-i m+1)]
\bigg\}\,.
\end{align}

\subsubsection{Symmetric tensor bases}
\label{app:tensor-basis-global}
The symmetric tensor bases in global coordinates can be decomposed using the dual basis one-forms $\{\text{d}\tau,\text{d}\varphi,\text{d}\psi\}$ via
\begin{equation}
\mathbf{W}^{(m\,h\,k)} = W^{(m\,h\,k)}_{ij}\,\text{d}x^i\otimes \text{d}x^j,\quad x\in\{\tau,\varphi,\psi\}\,.
\end{equation}
The tensor components are given by
\begin{align}
  W_{ij}^{(m\,h\,k)} \propto
  \begin{bmatrix}
    w^{(m\,h\,k)}_{\tau\tau} (\sin \psi )^{-2} &
    w^{(m\,h\,k)}_{\tau\varphi} (\sin \psi )^{-1}&
    w^{(m\,h\,k)}_{\tau\psi} (\sin \psi )^{-2}\\
    * &
    w^{(m\,h\,k)}_{\varphi\varphi} (\sin \psi )^{+0}&
    w^{(m\,h\,k)}_{\varphi\psi} (\sin \psi )^{-1}\\
    * &
    *&
    w^{(m\,h\,k)}_{\psi\psi} (\sin \psi )^{-2}
  \end{bmatrix}
  (\sin \psi )^{-h} e^{i [(h-k) \tau + m \varphi] +m \psi}\,,
\end{align}
where
\begin{align}
w_{\tau\tau}^{(m\,h\,0)}  &=\,  +\frac{1}{16}(c_1 e^{-2 i \psi }+4 c_1 e^{2 i \psi }-6 c_2 e^{-2 i \psi }+16 c_3 e^{2 i \psi }+8 c_5 e^{-2 i \psi }+16 c_6 e^{2 i \psi }+4 c_1-8 c_2+16 c_3+8 c_4)\,, \\ \nn
w_{\varphi\varphi}^{(m\,h\,0)} &=\,   c_1 \,, \\ \nn
w_{\psi\psi}^{(m\,h\,0)}   &=\,   +\frac{1}{16} (-8 c_4 +16 c_6 e^{2 i \psi }+c_1 e^{-2 i \psi }+2 c_2 e^{-2 i \psi }+8 c_5 e^{-2 i \psi })\,, \\ \nn
w_{\tau\varphi}^{(m\,h\,0)} &=\, -\frac{1}{4} \left(2 c_1 e^{ i \psi }+4 c_3 e^{ i \psi }+c_1 e^{ -i \psi }-2 c_2e^{ -i \psi }\right)\,, \\ \nn
w_{\varphi\psi}^{(m\,h\,0)} &=\,   +\frac{1}{4} \left(4 c_3 e^{ i \psi }+c_1 e^{ -i \psi }+2 c_2 e^{ -i \psi }\right)\,, \\ \nn
w_{\psi\tau}^{(m\,h\,0)} &=\,  -\frac{1}{16} \left(2 c_1 +4 c_2 +8 c_3 +8 c_3 e^{2 i \psi }+16 c_6 e^{2 i \psi }+c_1 e^{-2 i \psi }+2 c_2 e^{-2 i \psi }-8 c_5e^{-2 i \psi }\right)\,.
\end{align}

\clearpage{}

\section{Expressions of $\mathcal{D}^{(m,h)}_{A}[\mathbf{C}(u)]$ in Maxwell systems}
\label{app:G-function}

We have decomposed the differential operators
$\mathcal{D}^{(m,h)}_{A}[\mathbf{C}(u)],\,A\in\{T,\Phi,R,u\}$,
introduced in Sec.~\ref{sec:sep-vector}, by the
coefficients multiplying the 2nd, 1st, and
0th derivatives of the $C-$functions.  These
coefficients are tabulated here in Table~\ref{tab:maxwell}.
Expressions in this appendix can be computed using the companion
\textsc{Mathematica} notebook \texttt{Sep-met-pert-in-NHEK-Poinc.nb} \cite{NHEKsupplement}.

\begin{table}[htbp!]
\begin{equation*}
\begin{array}{c|cccc}
 \mathcal{D}_A & C_T''(u) & C_\Phi''(u) & C_R''(u) & C_u''(u) \\ 
 \noalign{\smallskip} \hline \hline \noalign{\smallskip}
 \mathcal{D}_T & \frac{1-u^2}{u^2+1} & 0 & 0 & 0 \\
 \mathcal{D}_\Phi  & 0 & \frac{1-u^2}{u^2+1} & 0 & 0 \\
  \mathcal{D}_R& 0 & 0 & \frac{1-u^2}{u^2+1} & 0 \\
 \mathcal{D}_u & 0 & 0 & 0 & 0 \\
   \noalign{\bigskip}
 \text{} & C_T'(u) & C_\Phi'(u) & C_R'(u) & C_u'(u) \\
 \noalign{\smallskip} \hline \hline \noalign{\smallskip}
 \mathcal{D}_T & -\frac{4 u}{\left(u^2+1\right)^2} & -\frac{2 u \left(u^2-3\right)}{\left(u^2+1\right)^2} & 0 & 0 \\
 \mathcal{D}_\Phi  & 0 & -\frac{2 u \left(u^2-1\right)}{\left(u^2+1\right)^2} & 0 & \frac{i m \left(u^2-1\right)}{u^2+1} \\
 \mathcal{D}_R & 0 & 0 & -\frac{4 u}{\left(u^2+1\right)^2} & \frac{h \left(u^2-1\right)}{u^2+1} \\
 \mathcal{D}_u & -\frac{i m}{u^2+1} & \frac{i m \left(u^4+6 u^2-3\right)}{4 \left(u^4-1\right)} & -\frac{h+1}{u^2+1} & 0 \\
   \noalign{\bigskip}
  \text{} & C_T(u) & C_\Phi(u) & C_R(u) & C_u(u) \\
 \noalign{\smallskip} \hline \hline \noalign{\smallskip}
  \multirow{2}{*}{$\mathcal{D}_T$} &
  \multicolumn{1}{l}{\frac{\left(u^4+6 u^2-3\right) m^2}{4(u^4-1)}+{}} &
  \multirow{2}{*}{$\frac{h \left(u^4+6 u^2-3\right)}{\left(u^2+1\right)^3}$} &
  \multirow{2}{*}{$-\frac{i m \left(u^4+6 u^2-3\right)}{\left(u^2+1\right)^3}$} &
  \multirow{2}{*}{$\frac{2 i m u \left(u^2-3\right)}{\left(u^2+1\right)^2}$} \\
 & \multicolumn{1}{r}{{}+\frac{(h+1) \left(-4 u^2+h \left(u^2+1\right)^2+4\right)}{\left(u^2+1\right)^3}} & & &\\
 \mathcal{D}_\Phi  & \frac{m^2 \left(u^2+1\right)^2-4 (h+1) \left(u^2-1\right)}{\left(u^2+1\right)^3} & \frac{h \left((h+1) u^4+2 (h+3) u^2+h-3\right)}{\left(u^2+1\right)^3} & -\frac{i m \left((h+1) u^4+2 (h+3) u^2+h-3\right)}{\left(u^2+1\right)^3} & \frac{2 i m u \left(u^2-1\right)}{\left(u^2+1\right)^2} \\
 \mathcal{D}_R & -\frac{i (h+1) m}{u^2+1} & \frac{i h m \left(u^4+6 u^2-3\right)}{4 \left(u^4-1\right)} & \frac{m^2 \left(u^4+6 u^2-3\right)}{4 \left(u^4-1\right)} & \frac{4 h u}{\left(u^2+1\right)^2} \\
 \mathcal{D}_u & 0 & 0 & 0 & \frac{4 \left(u^2-1\right) h^2+4 \left(u^2-1\right) h+m^2 \left(u^4+6 u^2-3\right)}{4 \left(u^4-1\right)} \\
\end{array}
\end{equation*}
\caption{%
  \label{tab:maxwell}
  The coefficient table that gives the expressions of
  $\mathcal{D}^{(m,h)}_{A}[\mathbf{C}(u)],\,A\in\{T,\Phi,R,u\}$ in
  Maxwell systems. Each row is labeled by $\mathcal{D}^{(m,h)}_{A}$,
  while each column is labeled by a $C$-function or its
  derivative. Each table component is the coefficient in front of the
  (derivative of) corresponding $C$-function in
  $\mathcal{D}^{(m,h)}_{A}[\mathbf{C}(u)]$. To recover
  $\mathcal{D}^{(m,h)}_{A}[\mathbf{C}(u)]$, one just multiplies each
  table component with its column label and then add up all those with
  the same row label $\mathcal{D}_{A}$.}
\end{table}

\clearpage{}

\section{Expressions of $\mathcal{D}^{(m,h)}_{AB}[\mathbf{C}(u)]$ in linearized Einstein equations}
\label{app:A-B-C-matrice}

The general second order differentiation $\mathcal{D}^{(m,h)}$ on the ten unknown $C$-functions, denoted
as $\mathcal{D}^{(m,h)}_{AB}[\mathbf{C}(u)]$, can be written compactly by putting all
$C$-functions together to form a vector $\mathbf{C}(u)$,
\begin{equation}
  \mathcal{D}^{(m,h)}_{AB}[\mathbf{C}(u)] = (\mathcal{A}_{AB} \partial^2_u 
    + \mathcal{B}_{AB}\partial_u + \mathcal{C}_{AB})\cdot
    \bigg(C_{TT}(u), \ldots, C_{\Phi u}(u)\bigg)^\text{T}.
\end{equation}
Here $\mathcal{A}_{AB}$, $\mathcal{B}_{AB}$, and $\mathcal{C}_{AB}$ are covectors whose
components are obtained by collecting coefficients in front of
$C$-functions. We further stack all the covectors $\mathcal{A}_{AB}$ to form a matrix, and similarly do for $\mathcal{B}_{AB}$ and $\mathcal{C}_{AB}$. We label the resulting
coefficient matrices as $\mathcal{A}, \mathcal{B}$, and
$\mathcal{C}$ respectively. They are given in Tables~\ref{tab:a-matrix-LEE}, \ref{tab:b-matrix-LEE}, \ref{tab:c-matrix-LEE-1}, \ref{tab:c-matrix-LEE-2}, and \ref{tab:c-matrix-LEE-3}.
They can also be computed using the companion \textsc{Mathematica}
notebook \texttt{Sep-met-pert-in-NHEK-Poinc.nb}, or read from the
precomputed expressions in \texttt{NHEK-precomputed.mx} \cite{NHEKsupplement}.

\begin{table}[htbp!]
\begin{equation*}
\begin{array}{c|cccccccccc}
\mathcal{D}_{AB} & C_{TT}''(u) & C_{T\Phi }''(u) & C_{\Phi \Phi }''(u) & C_{RR}''(u) & C_{Ru}''(u) & C_{uu}''(u) & C_{TR}''(u) & C_{Tu}''(u) & C_{\Phi R}''(u) & C_{\Phi u}''(u)
  \\
\noalign{\smallskip}
\hline \hline \noalign{\smallskip}
 \mathcal{D}_{TT} & -\frac{2 \left(u^2-1\right)^2}{\left(u^2+1\right)^3} & \frac{u^6+5 u^4-9 u^2+3}{\left(u^2+1\right)^3} & -\frac{\left(u^4+6 u^2-3\right)^2}{8 \left(u^2+1\right)^3} & \frac{u^6+5 u^4-9 u^2+3}{2 \left(u^2+1\right)^3} & 0 & 0 & 0 & 0 & 0 & 0 \\
 \mathcal{D}_{T \Phi} & -\frac{2 \left(u^2-1\right)^2}{\left(u^2+1\right)^3} & \frac{u^6+9 u^4-17 u^2+7}{2 \left(u^2+1\right)^3} & -\frac{u^6+5 u^4-9 u^2+3}{2 \left(u^2+1\right)^3} & \frac{2 \left(u^2-1\right)^2}{\left(u^2+1\right)^3} & 0 & 0 & 0 & 0 & 0 & 0 \\
 \mathcal{D}_{\Phi \Phi} & -\frac{2 \left(u^2-1\right)^2}{\left(u^2+1\right)^3} & \frac{4 \left(u^2-1\right)^2}{\left(u^2+1\right)^3} & -\frac{2 \left(u^2-1\right)^2}{\left(u^2+1\right)^3} & \frac{2 \left(u^2-1\right)^2}{\left(u^2+1\right)^3} & 0 & 0 & 0 & 0 & 0 & 0 \\
 \mathcal{D}_{RR} & \frac{u^2-1}{2 \left(u^2+1\right)} & \frac{1-u^2}{u^2+1} & \frac{u^4+6 u^2-3}{8 \left(u^2+1\right)} & 0 & 0 & 0 & 0 & 0 & 0 & 0 \\
 \mathcal{D}_{Ru} & 0 & 0 & 0 & 0 & 0 & 0 & 0 & 0 & 0 & 0 \\
 \mathcal{D}_{uu} & 0 & 0 & 0 & 0 & 0 & 0 & 0 & 0 & 0 & 0 \\
 \mathcal{D}_{TR} & 0 & 0 & 0 & 0 & 0 & 0 & \frac{u^2-1}{2 \left(u^2+1\right)} & 0 & 0 & 0 \\
 \mathcal{D}_{Tu} & 0 & 0 & 0 & 0 & 0 & 0 & 0 & 0 & 0 & 0 \\
 \mathcal{D}_{\Phi R} & 0 & 0 & 0 & 0 & 0 & 0 & 0 & 0 & \frac{u^2-1}{2 \left(u^2+1\right)} & 0 \\
 \mathcal{D}_{\Phi u} & 0 & 0 & 0 & 0 & 0 & 0 & 0 & 0 & 0 & 0 \\
\end{array}
\end{equation*}
\caption{$\mathcal{A}$ matrix.}
\label{tab:a-matrix-LEE}
\end{table}

\begin{table}
\begin{centering}
\begin{equation*}
\begin{array}{c|ccccc}
 \mathcal{D}_{AB} & C_{TT}'(u) & C_{T\Phi}'(u) & C_{\Phi \Phi }'(u) & C_{RR}'(u) & C_{Ru}'(u)
  \\
\noalign{\smallskip}
\hline \hline \noalign{\smallskip}
 \mathcal{D}_{TT} & \frac{2 u \left(u^4-4 u^2+3\right)}{\left(u^2+1\right)^4} & -\frac{4 u \left(u^2-3\right) \left(u^2-1\right)}{\left(u^2+1\right)^4} & -\frac{u \left(u^{10}+u^8-22 u^6+66 u^4-123 u^2+45\right)}{8 \left(u^2-1\right) \left(u^2+1\right)^4} & -\frac{u \left(u^6+u^4-13 u^2+3\right)}{\left(u^2+1\right)^4} & -\frac{h \left(u^2-1\right) \left(u^4+6 u^2-3\right)}{\left(u^2+1\right)^3} \\
 \mathcal{D}_{T\Phi} & \frac{2 u \left(u^4-4 u^2+3\right)}{\left(u^2+1\right)^4} & -\frac{4 u \left(u^4-4 u^2+3\right)}{\left(u^2+1\right)^4} & \frac{2 u \left(u^4-4 u^2+3\right)}{\left(u^2+1\right)^4} & -\frac{2 u \left(u^4-4 u^2+3\right)}{\left(u^2+1\right)^4} & -\frac{2 (2 h+1) \left(u^2-1\right)^2}{\left(u^2+1\right)^3} \\
 \mathcal{D}_{\Phi \Phi} & \frac{2 u \left(u^4-4 u^2+3\right)}{\left(u^2+1\right)^4} & -\frac{4 u \left(u^2-3\right) \left(u^2-1\right)}{\left(u^2+1\right)^4} & \frac{2 u \left(u^4-4 u^2+3\right)}{\left(u^2+1\right)^4} & -\frac{2 u \left(u^4-4 u^2+3\right)}{\left(u^2+1\right)^4} & -\frac{4 (h+1) \left(u^2-1\right)^2}{\left(u^2+1\right)^3} \\
 \mathcal{D}_{RR} & -\frac{u \left(u^2-3\right)}{\left(u^2+1\right)^2} & \frac{2 u \left(u^2-3\right)}{\left(u^2+1\right)^2} & \frac{u \left(u^2-3\right)^3}{8 \left(u^2-1\right) \left(u^2+1\right)^2} & 0 & \frac{u^2-1}{u^2+1} \\
 \mathcal{D}_{Ru} & \frac{h+1}{2 \left(u^2+1\right)} & -\frac{2 h+1}{2 \left(u^2+1\right)} & \frac{h \left(u^4+6 u^2-3\right)}{8 \left(u^4-1\right)} & \frac{1}{2 \left(u^2+1\right)} & 0 \\
 \mathcal{D}_{uu} & -\frac{u}{2 \left(u^4-1\right)} & \frac{u}{u^4-1} & -\frac{u \left(u^2+3\right)}{4 \left(u^4-1\right)} & \frac{u}{2 \left(u^4-1\right)} & 0 \\
 \mathcal{D}_{TR} & 0 & 0 & 0 & 0 & 0 \\
 \mathcal{D}_{Tu} & \frac{i m}{2 \left(u^2+1\right)} & -\frac{i m \left(u^4+6 u^2-3\right)}{8 \left(u^4-1\right)} & 0 & 0 & 0 \\
 \mathcal{D}_{\Phi R} & 0 & 0 & 0 & 0 & -\frac{i m \left(u^2-1\right)}{2 \left(u^2+1\right)} \\
 \mathcal{D}_{\Phi u} & \frac{i m}{2 \left(u^2+1\right)} & -\frac{i m}{2 \left(u^2+1\right)} & 0 & -\frac{i m}{2 \left(u^2+1\right)} & 0 \\
\noalign{\bigskip}
 \text{} & C_{uu}'(u) & C_{TR}'(u) & C_{Tu}'(u) & C_{\Phi R}'(u) & C_{\Phi u}'(u)
  \\
\noalign{\smallskip}
\hline \hline \noalign{\smallskip}
 \mathcal{D}_{TT} & \frac{u \left(u^2-1\right) \left(u^6+11 u^4-13 u^2+9\right)}{2 \left(u^2+1\right)^4} & 0 & -\frac{i m \left(u^2-1\right) \left(u^4+6 u^2-3\right)}{\left(u^2+1\right)^3} & 0 & \frac{i m \left(u^4+6 u^2-3\right)^2}{4 \left(u^2+1\right)^3} \\
 \mathcal{D}_{T \Phi} & \frac{4 u \left(u^2-1\right)^3}{\left(u^2+1\right)^4} & 0 & -\frac{i m \left(u^6+9 u^4-17 u^2+7\right)}{2 \left(u^2+1\right)^3} & 0 & \frac{i m \left(u^6+5 u^4-9 u^2+3\right)}{\left(u^2+1\right)^3} \\
 \mathcal{D}_{\Phi\Phi} & \frac{4 u \left(u^2-1\right)^3}{\left(u^2+1\right)^4} & 0 & -\frac{4 i m \left(u^2-1\right)^2}{\left(u^2+1\right)^3} & 0 & \frac{4 i m \left(u^2-1\right)^2}{\left(u^2+1\right)^3} \\
 \mathcal{D}_{RR} & -\frac{u \left(u^2-1\right)}{2 \left(u^2+1\right)} & 0 & \frac{i m \left(u^2-1\right)}{u^2+1} & 0 & -\frac{i m \left(u^4+6 u^2-3\right)}{4 \left(u^2+1\right)} \\
 \mathcal{D}_{Ru} & 0 & \frac{i m}{2 \left(u^2+1\right)} & 0 & -\frac{i m \left(u^4+6 u^2-3\right)}{8 \left(u^4-1\right)} & 0 \\
 \mathcal{D}_{uu} & 0 & 0 & 0 & 0 & 0 \\
 \mathcal{D}_{TR} & 0 & -\frac{u \left(u^2-3\right)}{\left(u^2+1\right)^2} & -\frac{\left(u^2-1\right) \left(-u^4-6 u^2+h \left(u^2+1\right)^2+3\right)}{2 \left(u^2+1\right)^3} & \frac{u \left(u^2-3\right)}{\left(u^2+1\right)^2} & -\frac{\left(u^2-1\right) \left(u^4+6 u^2-3\right)}{2 \left(u^2+1\right)^3} \\
 \mathcal{D}_{Tu} & 0 & \frac{h+2}{2 \left(u^2+1\right)} & 0 & 0 & 0 \\
 \mathcal{D}_{\Phi R} & 0 & 0 & \frac{2 \left(u^2-1\right)^2}{\left(u^2+1\right)^3} & 0 & -\frac{\left(u^2-1\right) \left(h \left(u^2+1\right)^2+4 \left(u^2-1\right)\right)}{2 \left(u^2+1\right)^3} \\
 \mathcal{D}_{\Phi u} & 0 & 0 & 0 & \frac{h+1}{2 \left(u^2+1\right)} & 0 \\
\end{array}
\end{equation*}
\caption{$\mathcal{B}$ matrix.}
\end{centering}
\label{tab:b-matrix-LEE}
\end{table}

\begin{table}
\begin{centering}
$
\begin{array}{c|cc}
 \mathcal{D}_{AB} & C_{TT}(u) & C_{T\Phi }(u) \\
\noalign{\smallskip}
\hline \hline \noalign{\smallskip}
 \mathcal{D}_{TT} & \frac{\left(u^2-1\right) \left(u^4+2 u^2+2 h^2 \left(u^2+1\right)^2+6 h \left(u^2+1\right)^2+9\right)}{\left(u^2+1\right)^5} & -\frac{u^8-28 u^6-42 u^4+36 u^2+2 h^2 \left(u^8+8 u^6+10 u^4-3\right)+3 h \left(u^8+8 u^6+10 u^4-3\right)-15}{2 \left(u^2+1\right)^5} \\
  \mathcal{D}_{T\Phi} & \frac{\left(u^2-1\right) \left(2 h^2 \left(u^2+1\right)^2+5 h \left(u^2+1\right)^2+8\right)}{\left(u^2+1\right)^5} & -\frac{h^2 \left(u^4+10 u^2-7\right) \left(u^2+1\right)^2+h \left(u^4+10 u^2-7\right) \left(u^2+1\right)^2-8 \left(3 u^6+4 u^4-5 u^2+2\right)}{2 \left(u^2+1\right)^5} \\
  \mathcal{D}_{\Phi \Phi } & \frac{2 \left(u^2-1\right) \left(h^2 \left(u^2+1\right)^2+2 h \left(u^2+1\right)^2+4\right)}{\left(u^2+1\right)^5} & -\frac{2 \left(u^2-1\right) \left(-3 u^4-6 u^2+2 h^2 \left(u^2+1\right)^2+h \left(u^2+1\right)^2+5\right)}{\left(u^2+1\right)^5} \\
  \mathcal{D}_{RR} & \frac{8 \left(u^6-8 u^4+9 u^2-2\right)-m^2 \left(u^2+1\right)^4}{8 \left(u^2-1\right) \left(u^2+1\right)^3} & \frac{-3 u^4+30 u^2+h \left(u^2+1\right)^2-7}{2 \left(u^2+1\right)^3} \\
  \mathcal{D}_{Ru} & -\frac{(h+1) u}{\left(u^2+1\right)^2} & \frac{2 u \left(u^2+h \left(u^2-1\right)-2\right)}{\left(u^2-1\right) \left(u^2+1\right)^2} \\
  \mathcal{D}_{uu} & \frac{m^2 \left(u^2+1\right)^4+4 h^2 \left(u^2-1\right) \left(u^2+1\right)^2+8 h \left(u^2-1\right) \left(u^2+1\right)^2+8 \left(u^6-u^4+u^2-1\right)}{8 \left(u^2-1\right)^2 \left(u^2+1\right)^3} & -\frac{3 u^4-2 u^2+2 h^2 \left(u^2+1\right)^2+3 h \left(u^2+1\right)^2+3}{2 \left(u^2-1\right) \left(u^2+1\right)^3} \\
  \mathcal{D}_{TR} & \frac{i m \left(u^4-2 u^2+2 h \left(u^2+1\right)^2+5\right)}{4 \left(u^2+1\right)^3} & -\frac{i m \left(u^4+6 u^2-3\right) \left(-u^4-6 u^2+h \left(u^2+1\right)^2+3\right)}{8 \left(u^2-1\right) \left(u^2+1\right)^3} \\
  \mathcal{D}_{Tu} & \frac{i m u}{2-2 u^4} & \frac{i m u \left(u^4+6 u^2-3\right)}{4 \left(u^2-1\right) \left(u^2+1\right)^2} \\
  \mathcal{D}_{\Phi R} & \frac{i m \left(u^4+h \left(u^2+1\right)^2+3\right)}{2 \left(u^2+1\right)^3} & -\frac{i m \left(-u^4-6 u^2+h \left(u^2+1\right)^2+3\right)}{2 \left(u^2+1\right)^3} \\
  \mathcal{D}_{\Phi u} & \frac{i m u}{2-2 u^4} & \frac{i m u}{\left(u^2+1\right)^2} \\
\noalign{\bigskip}
 \text{} & C_{\Phi R}(u) & C_{\Phi u}(u) \\
\noalign{\smallskip}
\hline \hline \noalign{\smallskip}
 \mathcal{D}_{TT} & -\frac{i h m \left(u^4+6 u^2-3\right)^2}{4 \left(u^2-1\right) \left(u^2+1\right)^3} & \frac{i m u \left(u^4+6 u^2-3\right)^2}{4 \left(u^2-1\right) \left(u^2+1\right)^3} \\
  \mathcal{D}_{T\Phi } & -\frac{i h m \left(u^4+6 u^2-3\right)}{\left(u^2+1\right)^3} & \frac{i m u \left(u^4+6 u^2-3\right)}{\left(u^2+1\right)^3} \\
  \mathcal{D}_{\Phi\Phi } & -\frac{4 i h m \left(u^2-1\right)}{\left(u^2+1\right)^3} & \frac{4 i m u \left(u^2-1\right)}{\left(u^2+1\right)^3} \\
  \mathcal{D}_{RR} & \frac{i m \left(u^4+6 u^2-3\right)}{4 \left(u^4-1\right)} & -\frac{i m u \left(u^6+3 u^4+19 u^2-15\right)}{4 \left(u^2-1\right) \left(u^2+1\right)^2} \\
  \mathcal{D}_{Ru} & \frac{i m u \left(u^4+6 u^2-3\right)}{4 \left(u^2-1\right) \left(u^2+1\right)^2} & -\frac{i h m \left(u^4+6 u^2-3\right)}{8 \left(u^4-1\right)} \\
  \mathcal{D}_{uu} & -\frac{i (h+1) m \left(u^4+6 u^2-3\right)}{4 \left(u^2-1\right)^2 \left(u^2+1\right)} & \frac{i m u \left(u^2+3\right)}{2 \left(u^4-1\right)} \\
  \mathcal{D}_{TR} & -\frac{u^4-12 u^2+3}{\left(u^2+1\right)^3} & \frac{2 u \left(u^4-14 u^2+9\right)}{\left(u^2+1\right)^4} \\
  \mathcal{D}_{Tu} & \frac{(h+2) u \left(u^2-3\right)}{\left(u^2-1\right) \left(u^2+1\right)^2} & -\frac{(h+2) \left(u^4+6 u^2-3\right)}{2 \left(u^2+1\right)^3} \\
  \mathcal{D}_{\Phi R} & \frac{6 u^2-2}{\left(u^2+1\right)^3} & -\frac{2 u \left(h \left(u^2+1\right)^2-2 \left(u^4-6 u^2+5\right)\right)}{\left(u^2+1\right)^4} \\
  \mathcal{D}_{\Phi u} & -\frac{2 (h+1) u}{\left(u^2-1\right) \left(u^2+1\right)^2} & -\frac{(h+1) \left(h \left(u^2+1\right)^2+4 \left(u^2-1\right)\right)}{2 \left(u^2+1\right)^3} \\
\end{array}
$
\end{centering}
\caption{Part I of $\mathcal{C}$ matrix.}
\label{tab:c-matrix-LEE-1}
\end{table}

\begin{table}
\begin{centering}
$
\begin{array}{c|cc}
\mathcal{D}_{AB} & C_{RR}(u) & C_{Ru}(u) \\
\noalign{\smallskip}
\hline \hline \noalign{\smallskip}
 \mathcal{D}_{TT} & \frac{8 \left(u^{10}-2 u^8-6 u^6-8 u^4+21 u^2-6\right)-m^2 \left(u^6+7 u^4+3 u^2-3\right)^2}{8 \left(u^2-1\right) \left(u^2+1\right)^5} & -\frac{4 u \left((2 h+3) u^4+2 (h-6) u^2+9\right)}{\left(u^2+1\right)^4} \\
  \mathcal{D}_{T\Phi} & \frac{-\left(u^8+8 u^6+10 u^4-3\right) m^2+2 h \left(u^2-1\right) \left(u^2+1\right)^2+8 \left(u^6+u^4-3 u^2+1\right)}{2 \left(u^2+1\right)^5} & -\frac{4 u \left(u^2-1\right) \left(h u^2+2 u^2+h-4\right)}{\left(u^2+1\right)^4} \\
  \mathcal{D}_{\Phi \Phi } & \frac{2 \left(u^2-1\right) \left(-m^2 \left(u^2+1\right)^2+h \left(u^2+1\right)^2+2 \left(u^4+2 u^2-1\right)\right)}{\left(u^2+1\right)^5} & -\frac{4 (h+1) u \left(u^2-1\right)}{\left(u^2+1\right)^3} \\
  \mathcal{D}_{RR} & \frac{u^2-1}{\left(u^2+1\right)^3} & \frac{4 u}{\left(u^2+1\right)^2} \\
  \mathcal{D}_{Ru} & -\frac{u}{\left(u^2+1\right)^2} & \frac{8 \left(u^6+3 u^4-5 u^2+1\right)-m^2 \left(u^8+8 u^6+10 u^4-3\right)}{8 \left(u^2-1\right) \left(u^2+1\right)^3} \\
  \mathcal{D}_{uu} & \frac{-\left(u^8+8 u^6+10 u^4-3\right) m^2+4 h \left(u^2-1\right) \left(u^2+1\right)^2+16 u^2 \left(u^2-1\right)}{8 \left(u^2-1\right)^2 \left(u^2+1\right)^3} & -\frac{(h+1) u}{u^4-1} \\
  \mathcal{D}_{TR} & \frac{i m \left(u^4+6 u^2-3\right)}{4 \left(u^2+1\right)^3} & -\frac{i m u \left(u^2-3\right)}{\left(u^2+1\right)^2} \\
  \mathcal{D}_{Tu} & -\frac{i m u \left(u^2-3\right)}{2 \left(u^2-1\right) \left(u^2+1\right)^2} & \frac{i m \left(u^4+6 u^2-3\right)}{2 \left(u^2+1\right)^3} \\
  \mathcal{D}_{\Phi R} & \frac{i m \left(u^4+4 u^2-1\right)}{2 \left(u^2+1\right)^3} & -\frac{i m u \left(u^2-1\right)}{\left(u^2+1\right)^2} \\
  \mathcal{D}_{\Phi u} & \frac{i m u}{2 \left(u^4-1\right)} & \frac{i m \left(u^4+6 u^2+h \left(u^2+1\right)^2-3\right)}{2 \left(u^2+1\right)^3} \\
\noalign{\bigskip}
 \text{} & C_{uu}(u) & C_{TR}(u) \\
\noalign{\smallskip}
\hline \hline \noalign{\smallskip}
 \mathcal{D}_{TT} & \frac{4 h^2 \left(u^6+5 u^4-9 u^2+3\right) \left(u^2+1\right)^2+m^2 \left(u^6+7 u^4+3 u^2-3\right)^2+8 \left(5 u^8+34 u^6-68 u^4+54 u^2-9\right)}{8 \left(u^2+1\right)^5} & \frac{i (2 h+3) m \left(u^4+6 u^2-3\right)}{2 \left(u^2+1\right)^3} \\
  \mathcal{D}_{T\Phi} & \frac{\left(u^2-1\right) \left(\left(u^8+8 u^6+10 u^4-3\right) m^2+4 h^2 \left(u^2-1\right) \left(u^2+1\right)^2+2 h \left(u^2-1\right) \left(u^2+1\right)^2+8 \left(u^6+8 u^4-11 u^2+2\right)\right)}{2 \left(u^2+1\right)^5} & \frac{i m \left(2 \left(u^4+8 u^2-5\right)+h \left(u^4+10 u^2-7\right)\right)}{2 \left(u^2+1\right)^3} \\
  \mathcal{D}_{\Phi \Phi } & \frac{2 \left(u^2-1\right)^2 \left(h^2 \left(u^2+1\right)^2+m^2 \left(u^2+1\right)^2+h \left(u^2+1\right)^2+2 \left(u^4+9 u^2-2\right)\right)}{\left(u^2+1\right)^5} & \frac{2 i (2 h+3) m \left(u^2-1\right)}{\left(u^2+1\right)^3} \\
  \mathcal{D}_{RR} & -\frac{\left(u^8+8 u^6+10 u^4-3\right) m^2+4 h \left(u^2-1\right) \left(u^2+1\right)^2+8 \left(u^4+4 u^2-1\right)}{8 \left(u^2+1\right)^3} & -\frac{i m}{2 \left(u^2+1\right)} \\
  \mathcal{D}_{Ru} & -\frac{h u}{2 \left(u^2+1\right)} & -\frac{i m u}{\left(u^2+1\right)^2} \\
  \mathcal{D}_{uu} & \frac{u^2 \left(u^2+3\right)}{\left(u^2+1\right)^3} & \frac{i (2 h+3) m}{2 \left(u^4-1\right)} \\
  \mathcal{D}_{TR} & \frac{i m \left(u^2-1\right) \left(u^4+6 u^2-3\right)}{4 \left(u^2+1\right)^3} & \frac{8 \left(u^6-7 u^4+7 u^2-1\right)-m^2 \left(u^8+8 u^6+10 u^4-3\right)}{8 \left(u^2-1\right) \left(u^2+1\right)^3} \\
  \mathcal{D}_{Tu} & -\frac{i m u \left(u^2-3\right)}{2 \left(u^2+1\right)^2} & -\frac{(h+2) u}{\left(u^2+1\right)^2} \\
  \mathcal{D}_{\Phi R} & \frac{i m \left(u^2-1\right) \left(h \left(u^2+1\right)^2+2 \left(u^2-1\right)\right)}{2 \left(u^2+1\right)^3} & -\frac{m^2}{2 \left(u^2+1\right)} \\
  \mathcal{D}_{\Phi u} & -\frac{i m u \left(u^2-1\right)}{\left(u^2+1\right)^2} & 0 \\
\end{array}
$
\end{centering}
\caption{Part II of $\mathcal{C}$ matrix.}
\label{tab:c-matrix-LEE-2}
\end{table}

\begin{table}
\begin{centering}
$
\begin{array}{c|c}
 \mathcal{D}_{AB} & C_{\Phi \Phi }(u) \\
\noalign{\smallskip}
\hline \hline \noalign{\smallskip}
 \mathcal{D}_{TT} & \frac{h^2 \left(u^2-1\right) \left(u^6+7 u^4+3 u^2-3\right)^2-2 \left(3 u^{12}+68 u^{10}-5 u^8-128 u^6+153 u^4-36 u^2+9\right)}{8 \left(u^2-1\right)^2 \left(u^2+1\right)^5} \\
  \mathcal{D}_{T\Phi} & -\frac{-2 \left(u^8+8 u^6+10 u^4-3\right) h^2+\left(u^8+8 u^6+10 u^4-3\right) h+4 \left(9 u^6+13 u^4-9 u^2+3\right)}{4 \left(u^2+1\right)^5} \\
  \mathcal{D}_{\Phi\Phi} & \frac{\left(u^2-1\right) \left(-3 u^4-6 u^2+2 h^2 \left(u^2+1\right)^2-2 h \left(u^2+1\right)^2+5\right)}{\left(u^2+1\right)^5} \\
  \mathcal{D}_{RR} & \frac{2 \left(7 u^8-30 u^6+72 u^4-42 u^2+9\right)-h \left(u^2+1\right)^2 \left(u^6+5 u^4-9 u^2+3\right)}{8 \left(u^2-1\right)^2 \left(u^2+1\right)^3} \\
  \mathcal{D}_{Ru} & -\frac{u \left(8 \left(u^4-4 u^2+3\right)+h \left(u^6+11 u^4-13 u^2+9\right)\right)}{8 \left(u^4-1\right)^2} \\
  \mathcal{D}_{uu} & \frac{\left(u^8+8 u^6+10 u^4-3\right) h^2+\left(u^8+8 u^6+10 u^4-3\right) h+2 \left(7 u^6+3 u^4+9 u^2-3\right)}{8 \left(u^2-1\right)^2 \left(u^2+1\right)^3} \\
  \mathcal{D}_{TR} & -\frac{i m \left(u^4+6 u^2-3\right)^2}{16 \left(u^2-1\right) \left(u^2+1\right)^3} \\
 \mathcal{D}_{Tu} & -\frac{i m u \left(u^6+3 u^4-21 u^2+9\right)}{8 \left(u^4-1\right)^2} \\
  \mathcal{D}_{\Phi R} & -\frac{i m \left(u^4+6 u^2-3\right)}{4 \left(u^2+1\right)^3} \\
  \mathcal{D}_{\Phi u} & -\frac{i m u \left(u^2-3\right)}{2 \left(u^2-1\right) \left(u^2+1\right)^2} \\
\noalign{\bigskip}
 \text{} & C_{Tu}(u) \\
\noalign{\smallskip}
\hline \hline \noalign{\smallskip}
 \mathcal{D}_{TT} & -\frac{2 i m u \left(u^2-1\right) \left(u^2+3\right)}{\left(u^2+1\right)^3} \\
 \mathcal{D}_{T\Phi} & -\frac{i m u \left(u^4+4 u^2-5\right)}{\left(u^2+1\right)^3} \\
  \mathcal{D}_{\Phi \Phi} & -\frac{4 i m u \left(u^2-1\right)}{\left(u^2+1\right)^3} \\
  \mathcal{D}_{RR} & \frac{4 i m u}{\left(u^2+1\right)^2} \\
  \mathcal{D}_{Ru} & \frac{i (h+1) m}{2 \left(u^2+1\right)} \\
  \mathcal{D}_{uu} & -\frac{i m u}{u^4-1} \\
  \mathcal{D}_{TR} & -\frac{2 u \left(u^4-14 u^2+h \left(u^2+1\right)^2+9\right)}{\left(u^2+1\right)^4} \\
  \mathcal{D}_{Tu} & -\frac{4 h^2 \left(u^2-1\right) \left(u^2+1\right)^2+4 h \left(u^6-3 u^4+7 u^2-5\right)+\left(u^4+6 u^2-3\right) \left(-8 u^2+m^2 \left(u^2+1\right)^2+8\right)}{8 \left(u^2-1\right) \left(u^2+1\right)^3} \\
  \mathcal{D}_{\Phi R} & -\frac{4 u \left(u^4-6 u^2+5\right)}{\left(u^2+1\right)^4} \\
  \mathcal{D}_{\Phi u} & \frac{-m^2 \left(u^2+1\right)^2+4 h \left(u^2-1\right)+4 \left(u^2-1\right)}{2 \left(u^2+1\right)^3} \\
\end{array}
$
\end{centering}
\caption{Part III of $\mathcal{C}$ matrix.}
\label{tab:c-matrix-LEE-3}
\end{table}

\clearpage{}

\end{widetext}

\end{appendix}

\bibliographystyle{apsrev4-1}
\bibliography{NHEK-and-stringy-corrections}

\end{document}